# Dynamics of B-cell repertoires and emergence of cross-reactive responses in COVID-19 patients with different disease severity

Zachary Montague[1*], Huibin Lv[2*], Jakub Otwinowski[3†], William S. DeWitt[4,5], Giulio Isacchini[3,6], Garrick K. Yip[2], Wilson W. Ng[2], Owen Tak-Yin Tsang[7], Meng Yuan[8], Hejun Liu[8], Ian A. Wilson[8,9], J. S. Malik Peiris[2], Nicholas C. Wu[10,11,12#], Armita Nourmohammad[1,3,5#], Chris Ka Pun Mok[2#]

1. Department of Physics, University of Washington, 3910 15th Ave Northeast, Seattle, WA 98195, USA
2. HKU-Pasteur Research Pole, School of Public Health, Li Ka Shing Faculty of Medicine, The University of Hong Kong, Hong Kong SAR, China.
3. Max Planck Institute for Dynamics and Self-organization, Am Faßberg 17, 37077 Göttingen, Germany
4. Department of Genome Sciences, University of Washington, 3720 15th Ave NE, Seattle, WA 98195, USA
5. Fred Hutchinson Cancer Research Center, 1100 Fairview Ave N, Seattle, WA 98109, USA
6. Laboratoire de physique de l'ecole normale supérieure (PSL University), CNRS, Sorbonne Université, and Université de Paris, 75005 Paris, France
7. Infectious Diseases Centre, Princess Margaret Hospital, Hospital Authority of Hong Kong
8. Department of Integrative Structural and Computational Biology, The Scripps Research Institute, La Jolla, CA 92037, USA
9. The Skaggs Institute for Chemical Biology, The Scripps Research Institute, La Jolla, CA 92037, USA
10. Department of Biochemistry, University of Illinois at Urbana-Champaign, Urbana, IL 61801, USA
11. Carl R. Woese Institute for Genomic Biology, University of Illinois at Urbana-Champaign, Urbana, IL 61801, USA
12. Center for Biophysics and Quantitative Biology, University of Illinois at Urbana-Champaign, Urbana, IL 61801, USA.

* equal contribution

† current address: Dyno Therapeutics, 1 Kendall Square, Building 1400E, Suite 202, Cambridge, MA 02139

# equal contribution and corresponding authors: Armita Nourmohammad: armita@uw.edu, Nicholas C. Wu: nicwu@illinois.edu, and Chris Ka Pun Mok: ch02mkp@hku.hk

## Abstract

COVID-19 patients show varying severity of the disease ranging from asymptomatic to requiring intensive care. Although a number of SARS-CoV-2 specific monoclonal antibodies have been identified, we still lack an understanding of the overall landscape of B-cell receptor (BCR) repertoires in COVID-19 patients. Here, we used high-throughput sequencing of bulk and plasma B-cells collected over multiple time points during infection to characterize signatures of B-cell response to SARS-CoV-2 in 19 patients. Using principled statistical approaches, we determined differential features of BCRs associated with different disease severity. We identified 38 significantly expanded clonal lineages shared among patients as candidates for specific responses to SARS-CoV-2. Using single-cell sequencing, we verified reactivity of BCRs shared among individuals to SARS-CoV-2 epitopes. Moreover, we identified natural emergence of a BCR with cross-reactivity to SARS-CoV-1 and SARS-CoV-2 in a number of patients. Our results provide important insights for development of rational therapies and vaccines against COVID-19.

## Introduction

The novel coronavirus SARS-CoV-2, which causes the severe respiratory disease COVID-19, has now spread to 216 countries and caused more than 120 million infections with a mortality rate around 2.2% (WHO, 2021). COVID-19 patients show varying disease severity ranging from asymptomatic to requiring intensive care. While epidemiological and clinical data report that many factors such as age, gender, genetic background, and preexisting conditions are associated with disease severity, host immunity against the virus infection is the crucial component of controlling disease progression (Ellinghaus et al., 2020; Guan et al., 2020; McKechnie and Blish, 2020; Vabret et al., 2020; Wu et al., 2020a). Shedding light on signatures of a protective immune response against SARS-CoV-2 infections can help elucidate the nature of COVID-19 and guide therapeutic developments as well as vaccine design and assessment.

Adaptive immunity is considered as one of the core protective mechanisms in humans against infectious diseases. A vast diversity of surface receptors on B- and T-cells enables us to recognize and counter new or repeated invasions from a multitude of pathogens (Janeway et al., 2005; Nielsen and Boyd, 2018). In particular, antibodies produced by B-cells can provide long-lasting protection against specific pathogens through neutralization or other antibody-mediated immune mechanisms (Janeway et al., 2005). During the early phase of an infection, antigens of a pathogen are recognized by a group of naïve B-cells, which then undergo affinity maturation in a germinal center through somatic hypermutation and selection. The B-cell receptors (BCRs) of mature B-cells can react strongly to infecting antigens, resulting in B-cell stimulation, clonal expansion, and ultimately secretion of high-affinity antibodies in the blood (Burnet, 1959, 1960; Cyster and Allen, 2019). The specificity of a BCR is determined by a number of features such as V-, (D-), or J-gene usage and length and sequence composition of the HCDR3 region. It has been found that SARS-CoV-2-specific IgG antibodies can be detected in plasma samples of COVID-19 patients starting from the first week post-symptom onset (Perera et al., 2020). These antibodies bind to different antigens including the spike protein and nucleoprotein as well as other structural or non-structural proteins (Hachim et al., 2020). In addition, multiple studies have isolated SARS-CoV-2-specific B-cells from COVID-19 patients and determined their germline origin (Barnes et al., 2020; Brouwer et al., 2020; Cao et al., 2020; Chi et al., 2020; Han et al., 2020; Hansen et al., 2020; Hurlburt et al., 2020; Ju et al., 2020; Kreer et al., 2020a; Kreye et al., 2020; Liu et al., 2020b; Noy-Porat et al., 2020; Robbiani et al., 2020; Rogers et al., 2020; Seydoux et al., 2020a, 2020b; Shi et al., 2020; Wu et al., 2020b; Yuan et al., 2020; Zost et al., 2020). However, we still lack a comprehensive view of patients' entire BCR repertoires during SARS-CoV-2 infections.

Antibody repertoire sequencing has advanced our understanding of the diversity of adaptive immune repertoires and their response to pathogens (Boyd et al., 2009; Georgiou et al., 2014; Kreer et al., 2020b; Robins, 2013). A few studies have performed BCR repertoire bulk sequencing to characterize the statistical signatures of the immune response to SARS-CoV-2 (Galson et al., 2020; Nielsen et al., 2020; Niu et al., 2020; Schultheiß et al., 2020). However, these studies have limited data on the dynamics of BCR repertoires, which could otherwise provide significant insight into responses specific to the infection. Moreover, they do not probe the composition of plasma B-cells during infection, which is the direct indicator of antibody production within an individual.

In this study, we have established a principled statistical approach to study the statistics and dynamics of bulk and plasma B-cell repertoires and to characterize the immune responses in 19 COVID-19 patients with different disease severities. By combining information from the statistics of sequence features in BCR repertoires, the expanding dynamics of clonal lineages during infection, and sharing of BCRs among COVID-19 patients, we identified 38 clonal lineages that are potential candidates for a response to SARS-CoV-2. Importantly, eight of these lineages contain BCRs from the plasma B-cell repertoire and, hence, are likely to have been secreting antibodies during infection. Moreover, using single-cell sequencing, we have verified the reactivity of BCRs shared among individuals to the epitopes of the receptor-binding domain (RBD) and the N-terminal domain (NTD) of SARS-CoV-2. Lastly, we identified cross-reactive responses to SARS-CoV-1 in some of the COVID-19 patients and a natural emergence of a previously isolated SARS-reactive antibody (Pinto et al., 2020) in three patients.

## Results

**Strong correlation between composition of bulk and plasma B-cell repertoires.** We obtained total RNA from the PBMC isolated from 19 patients infected with SARS-CoV-2 and three healthy individuals (see Methods, and Tables S1, S2 for details). To broaden our healthy control pool, we also incorporated into our analyses IgG B-cells from ten individuals in the Great Repertoire Project (GRP) (Briney et al., 2019). Sequence statistics for the first three biological replicates pooled together for each individual from the GRP are shown in Table S3 (see Methods). The patients showed different severities of symptoms, forming three categories of infected cohorts: two patients with mild symptoms, 12 patients with moderate symptoms, and five patients with severe symptoms. Specimens from all but one patient were collected over two or more time points during the course of the infection (Table S1). In addition to the bulk repertoire, we also isolated $CD38^{+}$ plasma B-cells from PBMC samples over at least two time points from seven patients in this cohort (six moderate, one severe) and from seven additional patients (two asymptomatic, three mild, two moderate) and three healthy individuals (Figure S1 and Table S4). The sampled time points for all patients in this study are indicated in Fig. 1 and Tables S1 and S4. IgG heavy chains of B-cell repertoires were sequenced by next-generation sequencing, and the statistics of the collected BCR read data from each sample are shown in Tables S1 and S2. Statistical models were applied to analyze the length of the HCDR3 region, IGHV- or IGHJ-gene usage, and expansion and sharing of specific clonal lineages (Fig. 1).

The bulk repertoire is a collection of all BCRs circulating in the blood, including receptors from naïve, memory, and plasma B-cells. Plasma B-cells are actively producing antibodies and their receptors are more likely to be engaged in responding to an ongoing infection. Interestingly, the abundance of B-cell clonal lineages in the bulk and the plasma are strongly correlated (Fig. S3A), with Pearson correlations ranging from 0.55 – 0.88 across patients and significance $\text{p} - \text{values} < 5 \times 10^{-8}$ across patients; correlations and p-values are given for each patient in Fig. S3. The significant correspondence between the bulk and plasma B-cell repertoires in Fig. 2 indicates that samples from the bulk, which cover a larger depth, are representative of functional immune responses, at least in the course of the infection.

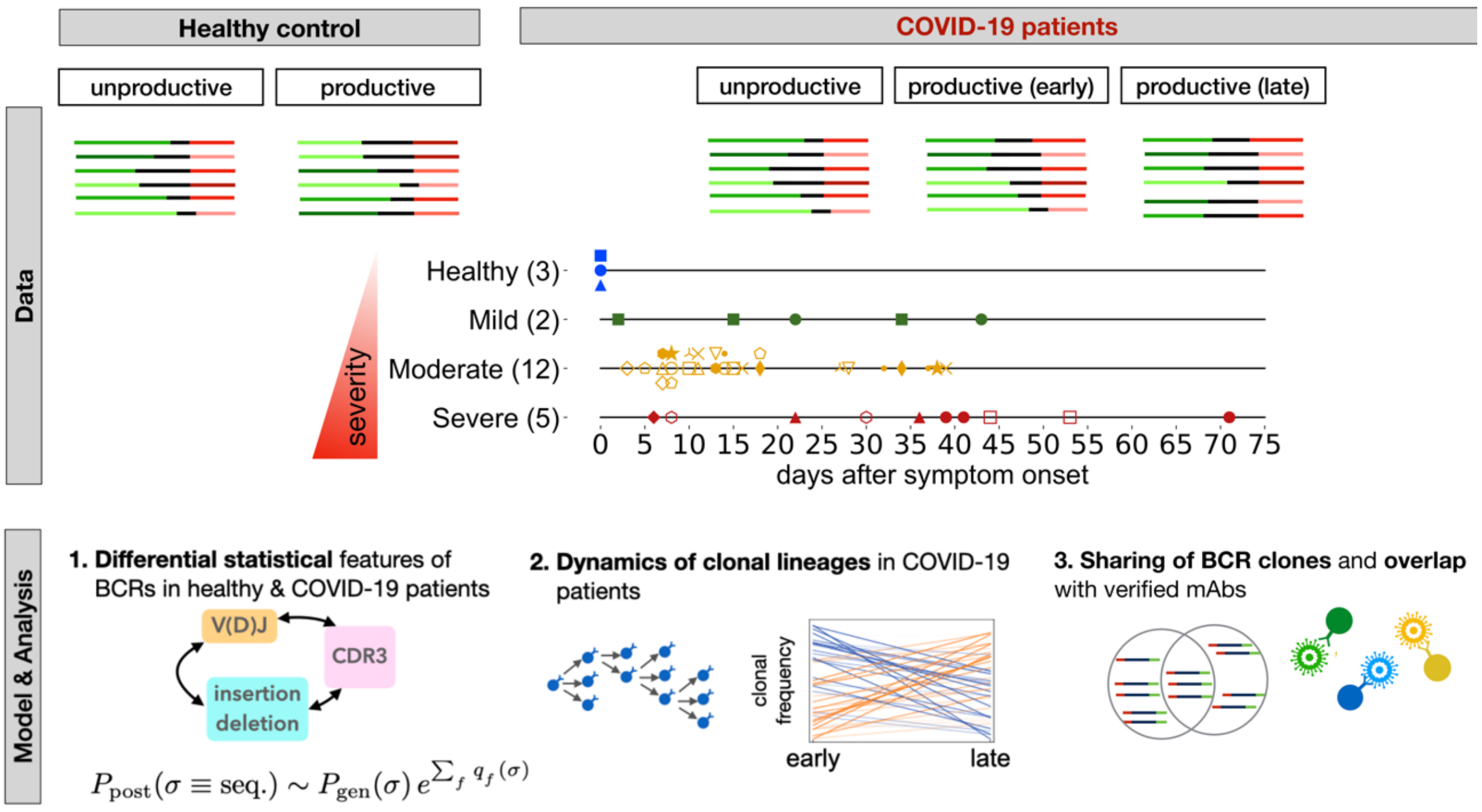

**Figure 1. Roadmap for analysis of BCR repertoires. Top:** We collected bulk blood IgG BCR samples from three healthy individuals and 19 COVID-19 patients where two patients had mild symptoms, 12 had moderate symptoms, and five had severe symptoms (different markers and colors); see Methods. We also collected $CD38^{+}$ plasma B-cells from PBMC samples of seven patients in this cohort (six moderate, one severe) and from seven additional patients (two asymptomatic, three mild, two moderate), and three healthy individuals (Fig. S2, Tables S1, S2). Samples were collected at different time points during infection (shown in center for bulk repertoires). We distinguished between productive receptors and unproductive receptors that had frameshifts due to V(D)J recombination. Line segments of varying lengths represent full V(D)J rearrangements (colors). In each patient, we constructed clonal lineages for productive and unproductive BCRs and inferred the naïve progenitor of the lineage (Methods). **Bottom:** 1. Using the set of unproductive inferred naïve BCRs, we inferred a model to characterize the null probability for generation of receptors $P_{\text{gen}}(\sigma)$(Marcou et al., 2018). We inferred a selection model (Sethna et al. 2020) to characterize the deviation from the null among inferred naïve productive BCRs, with the probability of entry to the periphery $P_{\text{post}}(\sigma)$ and selection factors $q_f(\sigma)$, dependent on receptor sequence features. 2. Based on temporal information of sampled BCRs, we identified clonal lineages that showed significant expansion during infection. 3. We identified progenitors of clonal lineages shared among individuals and assessed the significance of these sharing statistics based on the probabilities to find each receptor in the periphery. The shared expanding clonal lineages that contain plasma B-cells, are likely candidates for secreting responsive antibodies during infection. We verified reactivity of receptors to SARS-CoV-2 antigenic epitopes using sorted single-cell data. We also identified previously characterized monoclonal antibodies (mAbs) specific to SARS-CoV-2 and SARS-CoV-1.

**B-cell repertoires differ in receptor compositions across cohorts.** We aimed to investigate whether cohorts with different disease severities can be distinguished by molecular features of their B-cell repertoires. Since sequence features of immune receptors (e.g. HCDR3 length and V- or J-gene usage) are often associated with their binding specificity, we used statistical methods to compare these features both at the level of clonal lineages, including the inferred receptor sequence of lineage progenitors in the bulk (Figs. 2, S2) and in the plasma B-cell repertoires (Figs. S3), and also the unique sequences in the bulk (Figs. S2) and in the plasma B-cell repertoires (Figs. S3); see Tables S1, S3, S4 for detailed statistics of clonal lineages in each individual.

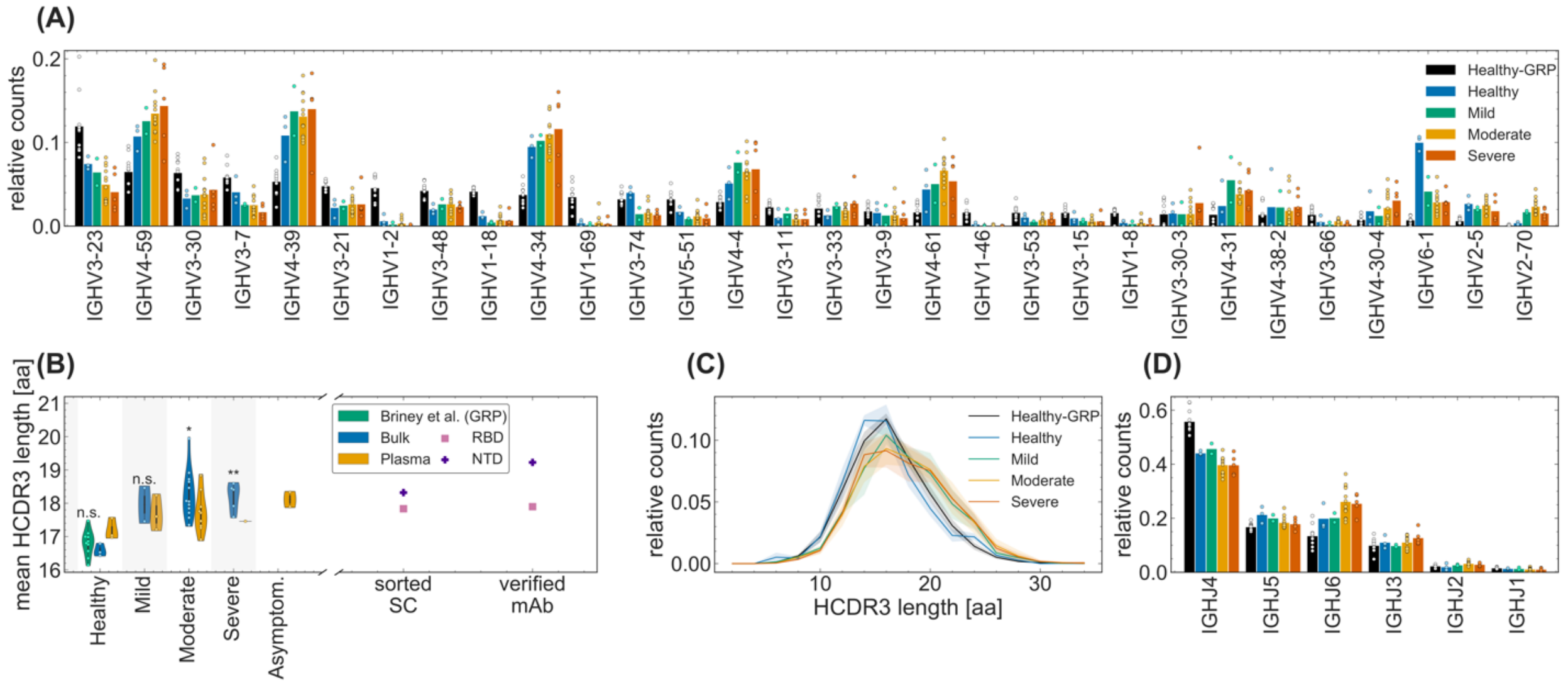

**Figure 2. Sequence features of immune receptors in the bulk repertoire across cohorts. (A)** The relative counts for IGHV-gene usage is shown for inferred naïve progenitors of clonal lineages in cohorts of healthy individuals and COVID-19 cohorts of patients exhibiting mild, moderate, and severe symptoms. The bars indicate the usage frequency averaged over individuals in each cohort, and dots indicate the variation in V-gene frequencies across individuals within each cohort. **(B, C)** Statistics of length of HCDR3 amino acid sequence is shown for different patients in each cohort. The violin plots in **(B)** show the mean HCDR3 length of each patient (dots) in a given cohort (color), with violin plot cut parameter set to 0.1. The mean HCDR3 lengths of the sorted single cells and verified monoclonal antibodies (axis) for RBD-reactive (pink squares) and NTD-reactive (purple pluses) receptors are shown on the right. Full lines in **(C)** show distributions averaged over individuals in each cohort (color), and shadings indicate regions containing one standard deviation of variation across individuals within a cohort. One-way ANOVA statistical tests were performed comparing the means HCDR3 of all the COVID-19 cohort and the healthy repertoires from Great Repertoire Project (GRP) dataset (Briney et al. 2019), with the healthy control from this study: Healthy-Mild: $F_{1,3} = 12.0$, p-value = 0.04; Healthy-Moderate: $F_{1,13} = 15.7$, p-value = 0.0016; Healthy-Severe: $F_{1,6} = 37.5$, p-value = 0.00087; Healthy-GRP: $F_{1,11} = 0.9$, p-value = 0.359. Significance cutoffs: n.s. $\mathrm{p-value} > 0.01$, * $\mathrm{p-value} \leq 0.01$, ** $\mathrm{p-value} < 0.001$. **(D)** The relative counts for IGHJ-gene usage is shown for inferred naïve progenitors of clonal lineages in cohorts of healthy individuals and COVID-19 cohorts of patients exhibiting mild, moderate, and severe symptoms. The bars indicate the usage frequency averaged over individuals in each cohort, and dots indicate the variation in J-gene frequencies across individuals within each cohort.

Lineage progenitors of IgG repertoires are closest to the ensemble of naïve receptors in the periphery. Features of lineage progenitors reflect receptor characteristics that are necessary for activating and forming a clonal lineage in response to an infection. In particular, the subset of lineages that contain plasma B-cell receptors can signal specific responses for antibody production against the infecting pathogen. Statistics of unique sequences in the bulk and the plasma B-cell repertoires, on the other hand, contain information about the size of the circulating lineages. Importantly, these statistical ensembles are relatively robust to PCR amplification biases that directly impact read abundances (see Methods for error correction and processing of reads).

IGHV genes cover a large part of pathogen-engaging regions of BCRs, including the three complementarity-determining regions HCDR1, HCDR2, and a portion of HCDR3. Therefore, we investigated if there are any differences in V-gene usage across cohorts, which may indicate preferences relevant for response to a particular pathogen. We found that the variation in V-gene usage among individuals within each cohort was far larger than differences among cohorts both in the bulk (Fig. 3A) and in the plasma B-cell repertoire (Fig. S3B). Data from unique sequences also indicated large background amplitudes due to vast differences in the sizes of lineages within a repertoire (Figs. S2A, S3E). Similarly, IGHJ-gene usage was also comparable across different cohorts for both bulk and plasma B-cell repertoires (Figs. 2D, S2C, and S3D,G). Moreover, we do not see a significant distinction in statistics of gene usage between the bulk and the plasma B cell repertoires (Figs. 2, S2 for bulk and Fig. S3 for plasma B-cells). Our results suggest that the SARS-CoV-2 V-gene specific responses are highly individualized at the repertoire level.

HCDR3 is part of the variable chain of B-cell receptors and is often a crucial region in determining specificity. Importantly, HCDR3 is highly variable in its sequence content and length due to insertion and deletion of sequence fragments at the VD and DJ junctions of the germline receptor. Therefore, differential characteristics of the HCDR3 sequence in BCR repertoires of different cohorts can signal preferences for sequence features specific to a class of antigens. We found that HCDR3s of lineages in COVID-19 patients with moderate and severe symptoms are significantly longer than in the healthy controls both from this study and from the GRP (Briney et al., 2019) (see Fig. 3B-C; One-way ANOVA statistics for differences in mean HCDR3 length: Healthy-Moderate: $F_{1,13} = 15.7$, p-value = $1.6 \times 10^{-3}$; Healthy-Severe: $F_{1,6} = 37.5$, p-value = $8.7 \times 10^{-4}$; GRP-Moderate: $F_{1,20} = 34.0$, p-value = $1.1 \times 10^{-5}$; GRP-Severe: $F_{1,13} = 41.5$, p-value = $2.2 \times 10^{-5}$). The difference between HCDR3 length in healthy individuals and patients with mild symptoms were less significant. These differences are also observed at the level of unique productive BCRs (Fig. S2B). These findings are consistent with previous reports of longer HCDR3 lengths in COVID-19 patients (Galson et al., 2020; Nielsen et al., 2020; Schultheiß et al., 2020). It should be noted that despite differences in experimental protocols, the HCDR3 length of the healthy cohort from this study and from GRP (Briney et al., 2019) are comparable to each other (Figs. 2B-C, S2B). In addition, we found no significant difference between the HCDR3 length of the unproductive BCR repertoires of healthy individuals and COVID-19 patients (Figs. S2E), which should reflect biases in the generation of receptors prior to functional selection. Taken together, these finding indicate that BCRs with a longer HCDR3 tend to be preferentially elicited in repertoires of individuals responding to SARS-CoV-2 infections. This preference seems to have a functional significance as longer HCDR3 is also observed among monoclonal antibodies (mAbs) specific to the receptor binding domain (RBD) and the N-terminal domain (NTD) of SARS-CoV-2 (Fig. 2B) which were identified in previous studies (Brouwer et al., 2020; Han et al., 2020; Hurlburt et al., 2020; Kreye et al., 2020; Pinto et al., 2020; Robbiani et al., 2020; Wu et al., 2020b; Zost et al., 2020).

**Differential selection on B-cell repertoires in response to SARS-CoV-2.** Longer HCDR3 sequences in COVID-19 patients can introduce more sequence diversity at the repertoire level. Quantifying sequence diversity of a B-cell repertoire can be very sensitive to the sampling depth in each individual. Despite progress in the quality of high-throughput repertoire sequencing techniques, sequenced BCRs still present a highly under-sampled view of the entire repertoire. To characterize the diversity of repertoires and the statistics of sequence features that make up this diversity, we inferred principled models of repertoire generation and selection for the entry of receptors into the periphery (Methods) (Elhanati et al., 2014; Marcou et al., 2018; Sethna et al., 2020). To do so, we first used data from unproductive lineage progenitors of B-cell receptors in the bulk repertoire to infer the highly non-uniform baseline model that characterizes the probability $P_{\text{gen}}(\sigma)$ to generate a given receptor sequence, dependent on its sequence features including the V-, D-, and J- gene choices and also the inserted and deleted sequences at the VD and DJ junctions (Elhanati et al., 2014; Marcou et al., 2018; Sethna et al., 2020) (Fig. 1 and Methods). The resulting model reflects the biased preferences in generating BCRs in the bone marrow by V(D)J recombination.

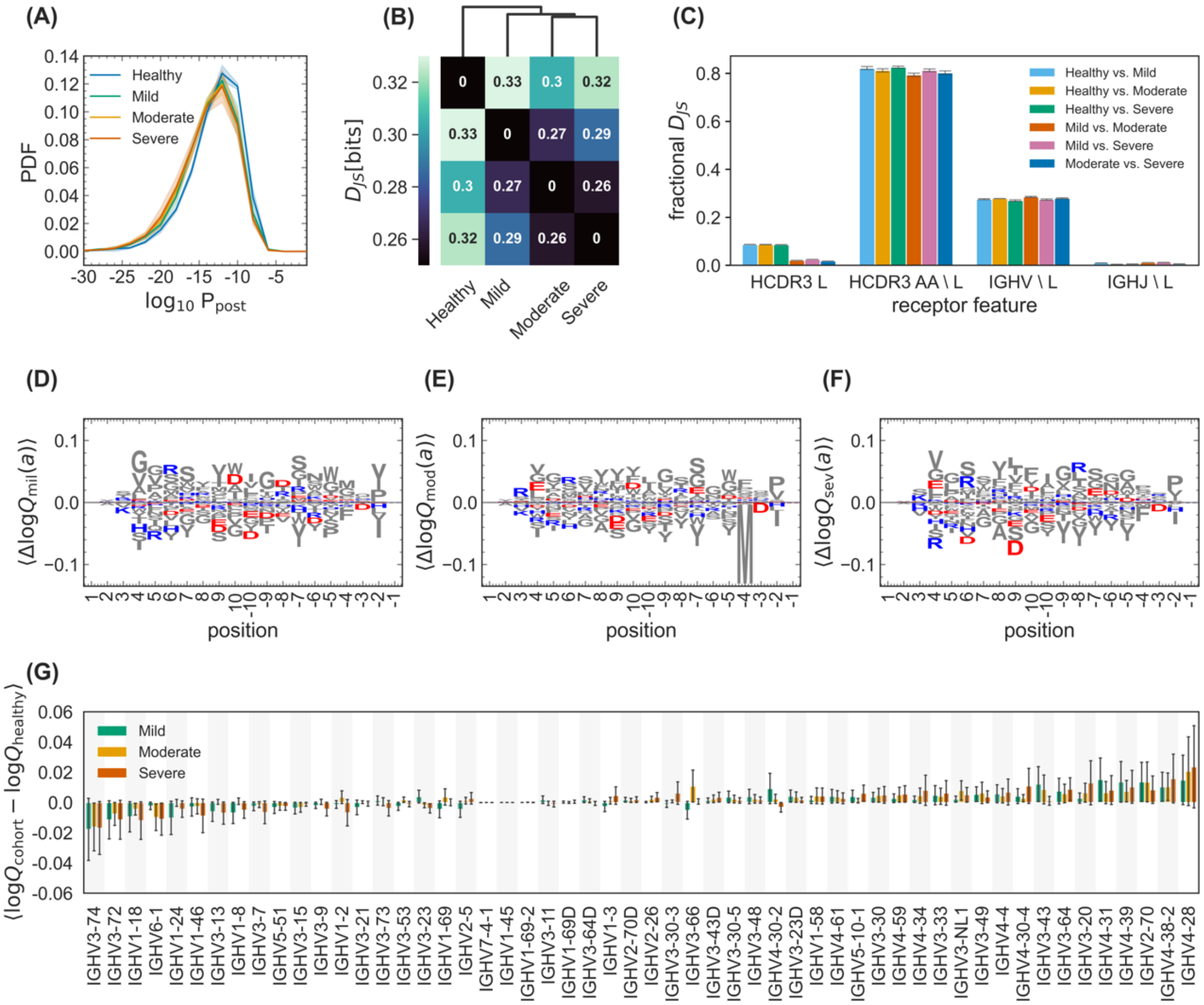

**Figure 3. Differential statistics of immune repertoires across cohorts. (A)** The distribution of the log-probability to observe a sequence $\sigma$ in the periphery $\log_{10} P_{\text{post}}(\sigma)$ is shown as a normalized probability density function (PDF) for inferred naïve progenitors of clonal lineages in cohorts of healthy individuals and the mild, moderate, and severe cohorts of COVID-19 patients. Full lines show distributions averaged over individuals in each cohort, and shadings indicate regions containing one standard deviation of variation among individuals within a cohort. **(B)** Clustering of cohorts based on their pairwise Jensen-Shannon divergences $D_{JS}$ as a measure of differential selection on cohorts is shown (Methods). **(C)** The bar graph shows how incorporating different features into a SONIA model contributes to the fractional Jensen-Shannon divergence between models trained on different cohorts. The error bars show the variations of these estimates over five independently inferred models (Methods). Logo plots show the expected differences in the log-selection factors for amino acid usage, $\langle \Delta \log Q_{\text{cohort}}(a) \rangle = \langle \log Q_{\text{cohort}}(a) - \log Q_{\text{healthy}}(a) \rangle$ for the **(D)** mild, **(E)** moderate, and **(F)** severe COVID-19 cohorts. The expectation values $\langle\cdot\rangle$ are evaluated on the mixture distribution $\frac{1}{2}\left(\mathrm{P}_{\text{post}}^{\text{cohort}} + \mathrm{P}_{\text{post}}^{\text{healthy}}\right)$. Positively charged amino acids (lysine, K; arginine, R; and histidine, H) are shown in blue while negatively charged amino acids (aspartate, D, and glutamate, E) are shown in red. All other amino acids are grey. Positions along the HCDR3 are shown up to 10 residues starting from the 3' (positive position values) and the 5' ends (negative position values). **(G)** The bar graph shows the average mean difference between the log-selection factors for IGHV-gene usage for the mild (green), moderate (yellow), and severe (red) COVID-19 cohorts, with the mean computed using the mixture distribution $\frac{1}{2}\left(\mathrm{P}_{\text{post}}^{\text{cohort}} + \mathrm{P}_{\text{post}}^{\text{healthy}}\right)$ and the average taken over the mean differences of 30 independently trained SONIA models for each cohort. Error bars show one standard deviation for the estimated mean, due to variations in the inferred SONIA models.

The functional, yet pathogen-naïve BCRs that enter the periphery experience selection through processes known as central tolerance (Janeway et al., 2005). In addition, the inferred progenitors of clonal lineages in the IgG repertoire have undergone antigen-dependent selection that led to expansion of their clonal lineages in response to an infection. These two levels of selection make sequence features of functional lineage progenitors distinct from the pool of unproductive BCRs that reflects biases of the generation process prior to any selection. In addition, differential selection on receptor features can be used to quantify a distance between repertoires of different cohorts that reflect their functional differences in responses to immune challenges (Isacchini et al., 2021).

To identify these distinguishing sequence features, we inferred a selection model for lineage progenitors (Methods). We characterized the probability to observe a clonal lineage ancestor in the periphery as $P_{\text{post}}(\sigma) \sim P_{\text{gen}}(\sigma)e^{\sum_{f:\text{features}} q_f(\sigma)}$, which deviates from the inferred generation probability of the receptor $P_{\text{gen}}(\sigma)$ by selection factors $q_f(\sigma)$ (Isacchini et al., 2020a, 2020b, 2021; Sethna et al., 2020). These selection factors $q_f(\sigma)$ depend on sequence features, including IGHV-gene and IGHJ-gene usages, HCDR3 length, and amino acid preferences at different positions in the HCDR3 (Methods) (Elhanati et al., 2014; Isacchini et al., 2020a, 2020b, 2021; Marcou et al., 2018; Sethna et al., 2020). Importantly, the inferred selection models are robust to the differences in the sample size of the repertoires, as long as enough data is available to train the models (Methods and Fig. S4C-F). As a result, selection models offer a robust approach to compare functional differences even between repertoires with widely different sample sizes, as is the case for our cohorts (Methods and Fig. S4C-F).

The distribution of the log-probability $\log_{10} P_{\text{post}}(\sigma)$ for the inferred progenitors of clonal lineages observed in individuals from different cohorts is shown in Fig. 3A. We find an overabundance of BCR lineages with progenitors that have a low probability of entering the periphery (i.e., a lower $P_{\text{post}}(\sigma)$) in COVID-19 patients compared to healthy individuals (Fig. 3A). A similar pattern is observed at the level of generation probability $P_{\text{gen}}(\sigma)$ for functional receptors in the healthy versus COVID-19 infected individuals (Fig. S4A). Notably, the inferred selection models from the GRP healthy repertoires are comparable to the healthy cohort in this study (Fig. S4B). Thus, the overabundance of rare receptors in COVID-19 patients is likely to be linked to functional responses associated with the stimulation of the repertoires against SARS-CoV-2.

We estimated the diversity of the repertoires in each cohort by evaluating the entropy of receptor sequences generated by the respective repertoire models (see Methods). In particular, diverse repertoires that contain B-cell lineages with rare receptors (i.e., those with a lower $P_{\text{post}}(\sigma)$), should have larger entropies. Based on this analysis, we find that immune repertoires are more diverse in COVID-19 patients compared to healthy individuals (Fig. 3A and Methods). Specifically, the entropy (i.e., diversity) of BCR bulk repertoires grows with severity of the disease, from $39.18$ bits in the healthy cohort to $40.81 \pm 0.03$ bits in the mild cohort, to $41.03 \pm 0.25$ bits in the moderate cohort, and to $41.32 \pm 0.11$ bits in the severe cohort (see Methods). The error bars indicate variations over different models inferred in each of the COVID-19 cohorts, from repertoires subsampled to the same size as the healthy control (Methods). As indicated in Fig. S4, the models inferred from subsampled repertoires are highly consistent within each cohort.

Selection factors $q_f(\sigma)$ determine the deviation in preferences for different sequence features of BCRs in each cohort, including their HCDR3 length and composition and IGHV-gene usages. A comparison of selection factors among cohorts can characterize their distinctive sequence features. To quantify the selection differences across cohorts, we evaluated the Jensen-Shannon divergence ($D_{\text{JS}}$) between repertoires of different cohorts, which measures the distance between the features of their receptor repertoire distributions (Isacchini et al., 2021) (Methods). Clustering of the cohorts based on their pairwise Jensen-Shannon divergences indicates that repertoires diverge with growing disease severity, and the COVID-19 cohorts are more similar with each other than with the healthy cohort (Fig. 3B, Methods).

The inferred selection models enabled us to quantify how different receptor features affect the pairwise divergence $D_{\text{JS}}$ of BCR repertoires (Methods). In particular, we found that HCDR3 length contributes the most to differences in receptor distributions between the healthy and COVID-19 cohorts (Fig. 3C), consistent with the significant differences in the HCDR3 length distributions shown in Fig. 2C. In addition, we found that the amino acid composition of HCDR3 is the second most distinguishing factor between repertoires (Fig. 3C), indicating that negatively charged amino acids are slightly suppressed at the center of HCDR3s in COVID-19 cohorts compared to healthy repertoires (Fig. 3D-F). The selection differences of IGHV- and IGHJ-gene usages between the healthy and the COVID-19 patients are insignificant (Figs. 3C,G), consistent with our previous analysis of lineage characteristics in Fig. 2A,D. Taken together, HCDR3 length and composition represents the molecular features that are most distinguishable at the repertoire level across different cohorts. Nonetheless, further work is necessary to understand the molecular underpinnings that may make these receptor features apt in response to a SARS-CoV-2 challenge.

**Expansion of BCR clonal lineages over time indicates responses to SARS-CoV-2.** Next, we examined the dynamics of BCR repertoires in the COVID-19 patients. The binding level (measured by $OD_{450}$ in ELISA assays) of both IgM and IgG antibodies against the receptor-binding domain (RBD) or N-terminal domain (NTD) of SARS-CoV-2 increased in most of the COVID-19 patients in our study over the course of their infection (Figs. 4A, S5). We expected that the increase of $OD_{450}$ binding level is associated with activation of specific B-cells, resulting in an increase in mRNA production of the corresponding BCRs. Detecting expansion of specific clonal lineages is challenging due to subsampling of the repertoires. In fact, only a limited overlap of BCR lineages was found if we simply compared the data between different time points or between replicates of a repertoire sampled at the same time point (Fig. S6). To identify expanding clonal lineages, we examined lineages only in patients whose plasma showed an increase in binding level ($OD_{450}$) to the RBD of SARS-CoV-2 and compared the sequence abundance of those lineages in the bulk repertoire that appeared in two or more time points (Figs. 4A, S6 and Methods). Using a hypothesis test with a false discovery rate of 7.5%, as determined by analyzing replicate data (Methods, Fig. S6), we detected significant expansion of clonal lineages of receptors harvested from the bulk repertoire within all investigated patients. The results reflect a dynamic repertoire in all patients, ranging from 5% to 15% of lineages with significant expansion and large changes in sequence abundances over time (Figs. 4, S6). The expanding lineages had comparable HCDR3 length to the rest of the repertoire in COVID-19 patients (Fig. S6). Moreover, we observed expanding lineages to show V-gene preferences comparable to those of previously identified antibodies against SARS-CoV-2 (RBD). This includes the abundance of IGHV4-59, IGHV4-39, IGHV3-23, IGHV3-53, IGH3-66, IGHV2-5, and IGHV2-70 (Brouwer et al., 2020; Ju et al., 2020; Pinto et al., 2020; Rogers et al., 2020). However, it should be noted that these preferences in V-gene usage among expanding lineages are comparable to the overall biases in V-gene usage within patients, and expanded lineages roughly make up 25% of lineages with a given V gene (Fig. 4C). Therefore, our results suggest that the overall response to SARS-CoV-2 is not driven by only a specific class of IGHV gene. We expect clonal expansions to reflect responses to SARS-CoV-2 during infection. Indeed, we observed that expanding lineages (based on the bulk data) show an over-representation of receptors harvested from plasma B-cells, which are likely to be associated with antibody-secreting B-cells (Fig. 4D and Methods); patient-specific significance p-values are reported in the caption of Fig. 4D.

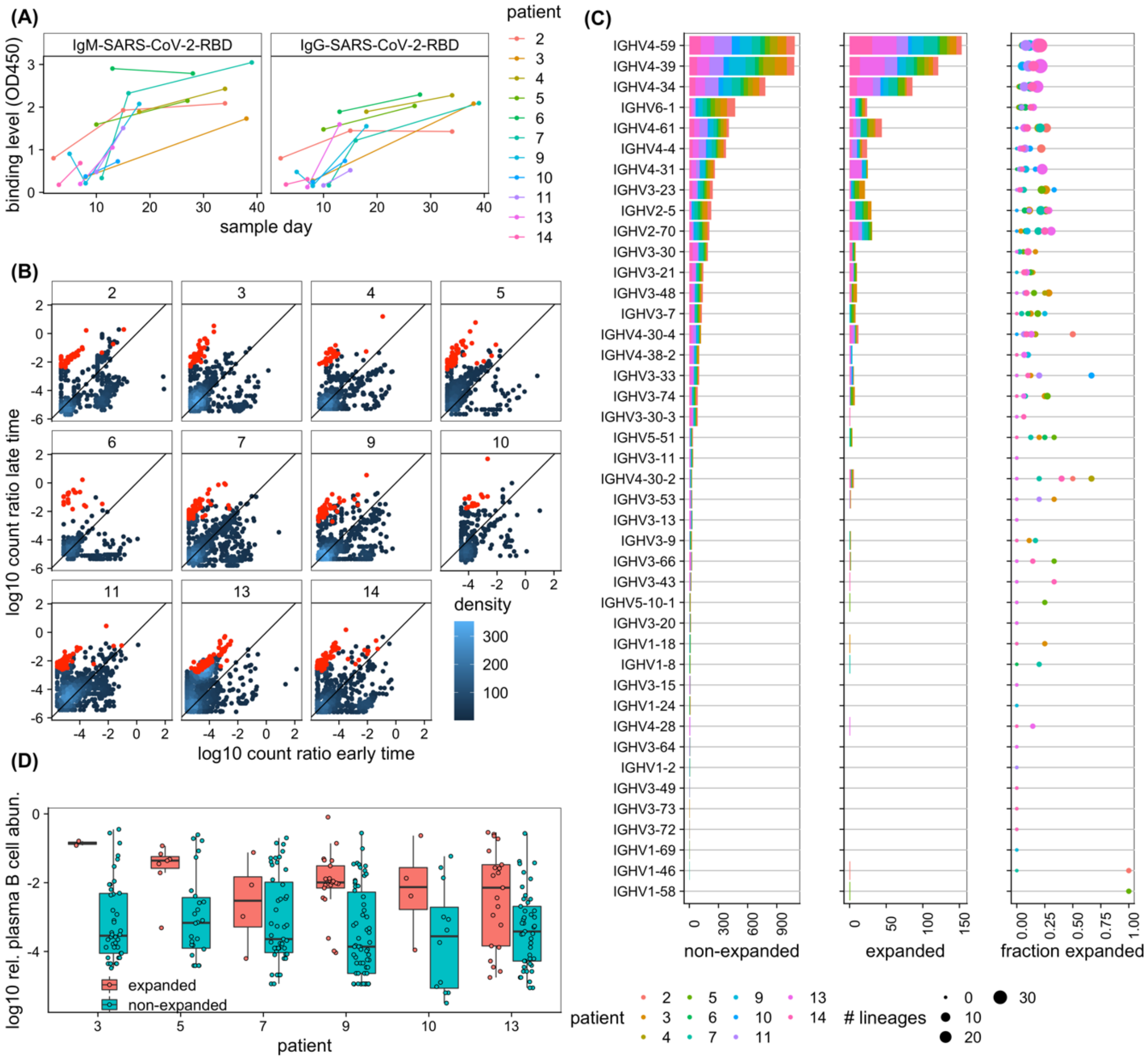

**Figure 4. Dynamics of BCR repertoires during infection. (A)** The binding level (measured by $OD_{450}$ in ELISA assay) of the IgM (left) and IgG (right) repertoires to SARS-CoV-2 (RBD) epitopes increases over time in most individuals. **(B)** The log-ratio of BCR (mRNA) abundance at late time versus early time is shown for all clonal lineages that are present at least in two time points (see Methods). Each panel shows dynamics of lineages for a given individual, as indicated in the label. The analysis is shown in individuals for whom the binding level ($OD_{450}$) of the IgG repertoire increases over time (shown in (**A**)). The count density indicates the number of lineages at each point. Lineages that show a significant expansion over time are indicated in red (see Methods for estimation of associated p-values). **(C)** IGHV-gene usage of lineages is shown for non-expanded (left) and expanded (middle) lineages in all individuals (colors). The right panel shows, for each patient (colors), the fraction of expanded lineages with a given IGHV gene as the number of expanded lineages divided by the total number of lineages with that given IGHV gene. The size of the circles indicates the total number of lineages in each category. **(D)** Boxplot of $\log_{10}$ relative read abundance in the plasma B-cell (Methods) are shown for expanding (red) and non-expanding (cyan) lineages that contain reads from the plasma B-cell in different patients. Receptors from the plasma B-cell are significantly more abundant in expanding lineages in a number of patients based on the ANOVA test statistics: patient 3: $F_{1,42} = 5.4$, p-value = 0.02; patient 5: $F_{1,31} = 0.5$, p-value = 0.5; patient 7: $F_{1,49} = 0.01$, p-value = 0.91; patient 9: $F_{1,42} = 4.1$, p-value = 0.04; patient 10: $F_{1,42} = 2.9$, p-value = 0.1; patient 13: $F_{1,64} = 7.7$, p-value = 0.007.

**Sharing of BCRs among individuals.** Despite the vast diversity of BCRs, we observe a substantial number of identical progenitors of BCR clonal lineages among COVID-19 patients (Fig. 5) and among healthy individuals from our dataset and from the GRP (Fig. S7). Previous work has also identified sharing of BCRs among COVID-19 patients, which was interpreted by the authors as evidence for large-scale convergence of immune responses (Galson et al., 2020; Nielsen et al., 2020; Schultheiß et al., 2020). Although BCR sharing can be due to convergent response to common antigens, it can also arise from convergent recombination leading to the same receptor sequence (Elhanati et al., 2018; Pogorelyy et al., 2018a) or simply from experimental biases. Therefore, it is imperative to formulate a null statistical model to identify the outliers among shared BCRs as candidates for common responses to antigens. Convergent recombination defines a null expectation for the amount of sharing within a cohort based on only the underlying biases for receptor generation within a repertoire (Elhanati et al., 2018; Pogorelyy et al., 2018a) (Methods). Intuitively, sharing is more likely among commonly generated receptors (i.e., with a high $P_{\text{post}}(\sigma)$) and within cohorts with larger sampling (Methods). Importantly, rare receptors (i.e., with a low $P_{\text{post}}(\sigma)$) that are shared among individuals in a common disease group can signal commonality in function and a response to a common antigen, as previously observed for TCRs in response to a yellow fever vaccine (Pogorelyy et al., 2018b) and CMV and diabetes (Pogorelyy et al., 2018a).

We used the receptors' probabilities $P_{\text{post}}(\sigma)$ to assess the significance of sharing by identifying a probabilistic threshold to limit the shared outliers both among the COVID-19 patients (dashed line in Fig. 5) and the healthy individuals (dashed lines in Fig. S7). Out of a total of 40,128 (unique) progenitors of clonal lineages reconstructed from the pooled bulk+plasma B-cell repertoires (Fig. 5A, Tables S1, S4), we found 10,146 progenitors to be shared among at least two individuals, and 761 of these lineages contained receptors found in the plasma B-cell of at least one individual. 167 of the 10,146 lineage progenitors were classified as rare, having a probability of occurrence below the indicated threshold (dashed line) in Fig. 5B, with 30 of them containing receptors harvested from plasma B-cell, indicating a significant over-abundance of plasma B-cells among the rare, shared receptors (p-value=$7.2 \times 10^{-6}$). Moreover, we found that 615 lineages shared a common sequence ancestor in at least two individuals and have expanded in at least one of the individuals (Fig. 5C-D). 38 of these shared, expanding lineages stemmed from rare naïve progenitors (below the dashed line in Fig. 5B, D), eight of which contain receptors found in the plasma B-cell of at least one individual. The over-abundance of plasma B-cell receptors in the rare, shared expanding lineages is significant (p-value=0.04). The sharing of these rare, expanding BCRs among COVID-19 patients, with an over-abundance of receptors associated with antibody production in the plasma B-cell data, indicates a potentially convergent response to SARS-CoV-2; these receptors are listed in Table S5.

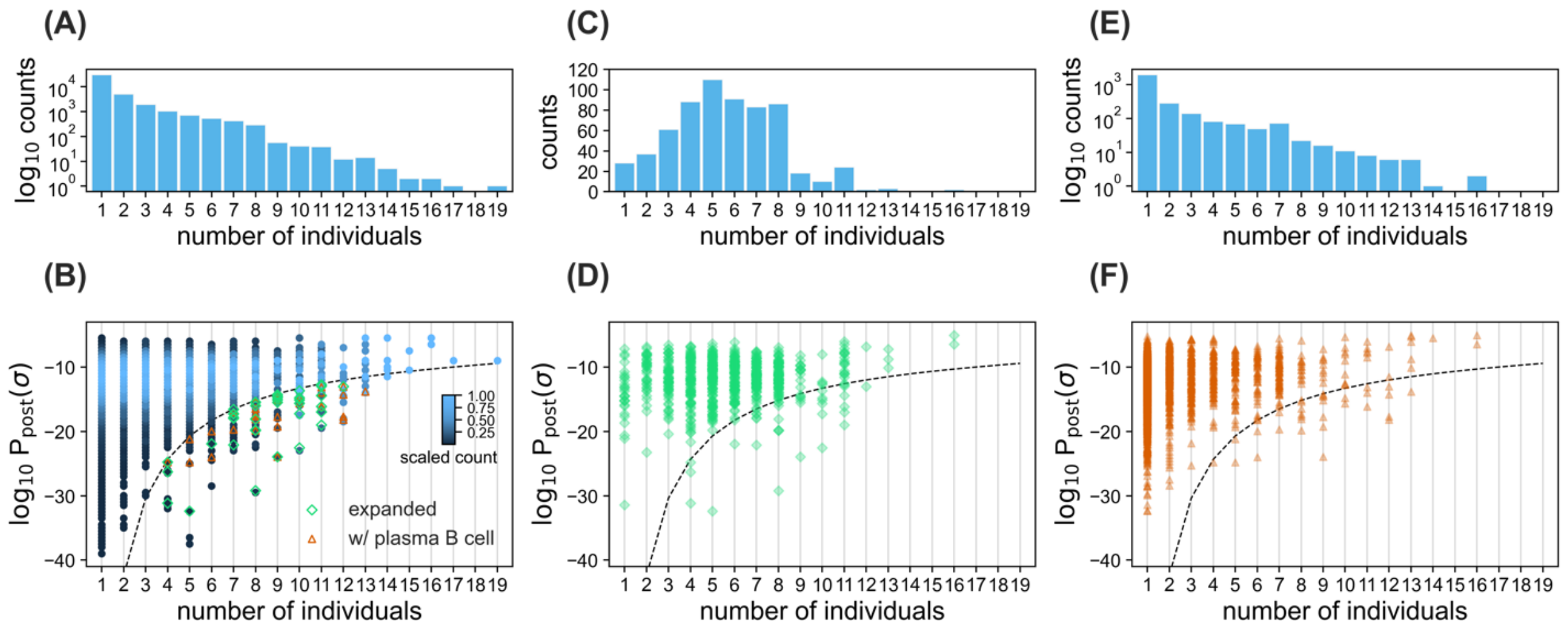

**Figure 5. Sharing of BCRs among patients.** **(A)** The histogram shows the number of clonal lineages that share a common progenitor in a given number of individuals, indicated on the horizontal axis. **(B)** The density plot shows the distribution of $\log_{10} P_{\text{post}}$ for progenitors of clonal lineages shared in a given number of individuals, indicated on the horizontal axis. Histogram bin size is 0.5. The scaling of sequence counts sets the maximum of the density in each column to one. Sharing of rare lineages with $\log_{10} P_{\text{post}}$ below the dashed line is statistically significant (Methods). Green diamonds indicate clonal lineages below the dashed line with significant expansion in at least one of the individuals. Orange triangles indicate clonal lineages below the dashed line that contain reads from the plasma B cell repertoire in at least one of the individuals. **(C, E)** The histograms show the number of clonal lineages that share a common progenitor in a given number individuals, which have significantly expanded during infection in at least one of the individuals (C), or contained reads from the plasma B-cell repertoire in at least one of the individuals (E). **(D, F)** The scatter plots with transparent overlapping markers show $\log_{10} P_{\text{post}}$ for progenitors of clonal lineages shared in a given number individuals that have expanded (D), or contain reads from the plasma B cell repertoire (E), in at least one individual. The dashed line is similar to (B).

Interestingly, we found that 24% of receptors in the 38 rare shared, expanding lineages contain multiple cysteines in their HCDR3s, in contrast to only 10% of the receptors in the whole repertoire. Such sequence patterns with cysteine pairs in the HCDR3 have been associated with stabilization of the HCDR3 loop by forming disulfide bonds with particular patterns and spacings of the cysteines (Lee et al., 2014; Prabakaran and Chowdhury, 2020). Disulfide bonds in the HCDR3 can decrease the conformational flexibility of the loop, thus decreasing the entropic cost of binding to improve the affinity of the receptor (Almagro et al., 2012). The significantly larger fraction of multi-cysteine HCDR3s among the candidate SARS-CoV-2 responsive receptors (p-value = 0.013 based on binomial sampling) indicates an underlying molecular mechanism for developing a potent response to SARS-CoV-2.

**Presence of SARS-CoV-2 and SARS-CoV-1 specific neutralizing antibodies within repertoires.**

To further investigate the functional response in the repertoire of COVID-19 patients, we performed single-cell sequencing on pooled samples from all patients, sorted for reactivity to RBD or NTD epitopes of SARS-CoV-2 (Methods). This analysis suggests that about 0.2% of these single cells are RBD-reactive as opposed to only 0.02% that are NTD-reactive (Fig. S1). This inferred fraction of reactive antibodies is consistent with previous estimates (Kreer et al., 2020a).

Next, we characterized the sequence features of RBD- and NTD-sorted antibodies. The IGHV-gene usage of these reactive receptors is shown in Fig. 6 and is compared to gene usage in monoclonal antibodies (mAbs) identified in previous studies (Brouwer et al., 2020; Han et al., 2020; Hurlburt et al., 2020; Kreye et al., 2020; Pinto et al., 2020; Robbiani et al., 2020; Wu et al., 2020b; Zost et al., 2020). Despite the broad range of IGHV-gene usages associated with epitope reactivity, sorted single-cell data show common IGHV-gene preferences to that of the previously identified mAbs against SARS-CoV-2 epitopes. This includes an abundance of IGHV1-69, IGHV4-59, IGHV3-30-3, IGHV3-33, IGHV1-18, IGHV5-51, and IGHV1-46 against RBD, and IGHV3-23, IGHV4-59, IGHV4-39, IGHV3-21, and IGHV3-48 against NTD (Fig. 7A). Similarly, we observe consistent biases in V- and J- gene usages of the $\kappa$ and $\lambda$ light chains for the sorted single-cell data and the verified mABs (Fig. S8). Moreover, the HCDR3 length distributions of the sorted single-cell data are comparable to those of the verified mABs (Fig. S8). The average length of the HCDR3 for both the verified mAbs and the sorted single-cell receptors are comparable to that of bulk repertoires from COVID-19 patients, which is significantly longer than that of healthy individuals (Fig. 2B).

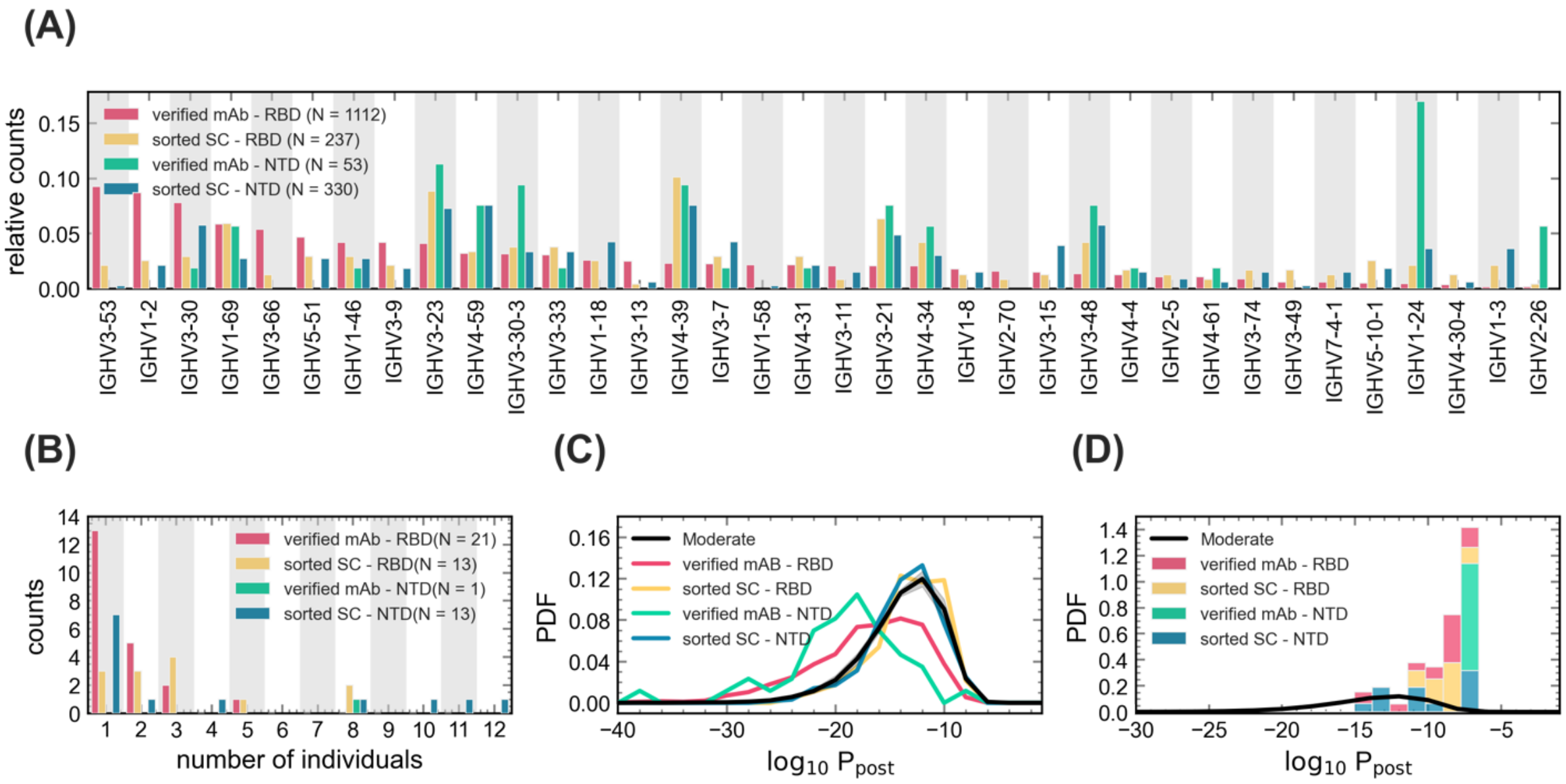

**Figure 6: Statistics of BCRs reactive to RBD and NTD epitopes. (A)** The relative counts for IGHV-gene usage is shown for known mAbs (Table S8) reactive to RBD (pink) and NTD (green) epitopes of SARS-CoV-2 and for receptors obtained from single cell sequencing of the pooled sample from all patients (Methods), sorted for RBD (yellow) and NTD (blue) epitopes. **(B)** The histogram shows the number of NTD-sorted receptors from single cell sequencing (Table S6) and RBD- and NTD-specific verified mAbs (Table S7) found in the bulk+plasma B-cell repertoires of a given number of individuals (Methods), indicated on the horizontal axis. **(C)** The distribution of the log-probability to observe a sequence $\sigma$ in the periphery $\log_{10} P_{\text{post}}(\sigma)$ is shown as a normalized probability density function (PDF) for inferred naïve progenitors of known RBD- and NTD-specific mAb and for RBD- and NTD-sorted receptors from single cell sequencing. $P_{\text{post}}(\sigma)$ values were evaluated based on the repertoire model created from patients with moderate symptoms. The corresponding $\log_{10} P_{\text{post}}$ distribution for bulk repertoires of the moderate cohort (similar to Fig. 3A) is shown in black as a reference. **(D)** Similar to (C) but restricted to receptors that are found in the bulk+ plasma B-cell repertoire of at least one patient in the cohort (Tables S6, S7). Colors are consistent between panels and the number of samples used to evaluate the statistics in each panel is indicated in the legend.

To characterize how SARS-CoV-2 reactive receptors make up the patients' repertoires, we mapped the heavy chain receptors from the sorted single-cell data onto BCR lineages constructed from the bulk+plasma B-cell data in the COVID-19 patients (Methods, Table S6). We found that 13 (from 237) RBD-sorted and 13 (from 330) NTD-sorted antibodies from the single-cell data matched receptor lineages in at least one individual (Fig. 6B). Interestingly, we found a broad sharing of these antibodies with 10 RBD- and 6 NTD-sorted single cells present in at least two patients (Fig. 6B).

In repertoires of the COVID-19 patients, we found that several HCDR3s matched with SARS-CoV-2-specific mAbs that were previously isolated in other studies (Brouwer et al., 2020; Han et al., 2020; Hurlburt et al., 2020; Kreye et al., 2020; Pinto et al., 2020; Robbiani et al., 2020; Wu et al., 2020b; Zost et al., 2020). Specifically, a total of 20 mAb families specific to SARS-CoV-2 epitopes were found to be close in sequence to HCDR3s in our data (with up to one amino acid difference), among which are 14 RBD-specific, one NTD-specific, and five S1-specific (reactive to either RBD or NTD) mAbs (Fig. 7B, Table S7). Interestingly, nine of these mAbs are shared among at least two individuals, and the NTD-specific antibody is found in eight individuals (Fig. 7B).

In addition, we found that two patients had exact HCDR3 matches to a previously identified antibody, S304, that has cross-reactivity to SARS-CoV-1 and SARS-CoV-2 (Pinto et al., 2020). We also observed in one patient an HCDR3 with only one amino acid difference to this antibody (Table S7). Importantly, the plasma in these patients showed a substantial binding level ($OD_{450}$) to SARS-CoV-1 (Fig. S5), which indicates a possibility of cross-reactive antibody responses to SARS-CoV-1 and SARS-CoV-2.

We also investigated the matches between the RBD- and NTD-sorted single-cell receptors with the verified mAbs from previous studies (Brouwer et al., 2020; Han et al., 2020; Hurlburt et al., 2020; Kreye et al., 2020; Pinto et al., 2020; Robbiani et al., 2020; Wu et al., 2020b; Zost et al., 2020). Although we found no matches between the heavy chain CDR3 of sorted single-cell receptors and the verified mAbs, we found a large number of matches between the $\kappa$ and $\lambda$ light chain CDR3s of the sets (Fig. S8). Notably, 59 of 142 $IG_\kappa$ and 47 of 110 $IG_\lambda$ from the RBD-reactive single cells, and 1 of 202 $IG_\kappa$, and 22 of 155 $IG_\lambda$ from the NTD-reactive single cells matched to light chain CDR3s of mAbs in those respective subsets (Fig. S8). Given the low sequence diversity of light chain receptors, it remains to be seen as to whether these matches between the light chain mAbs and sorted single-cell data are statistically significant—a question that would require modeling the generation and selection of the light chain receptors' repertoire.

Lastly, we observed that the previously verified mAbs have a lower probability $P_{\text{post}}(\sigma)$ of generation and entry to the periphery compared to the overall repertoire (Fig. 6C). This is in part expected since the selection models used to evaluate these probabilities were trained on different repertoires than those from which the mAbs were originally harvested. Consistently, the evaluated probabilities for the sorted single sorted receptors are within the range for the bulk repertoire (Fig. 6C), as the two datasets were derived from the same cohort. It should also be noted that all of the verified mAbs and the sorted receptors from the single-cell data that we can match to the patients' repertoires have a relatively high probability $P_{\text{post}}(\sigma)$ (Fig. 6D). This is not surprising as it is very unlikely to observe rare BCRs (with small $P_{\text{post}}(\sigma)$) to be shared in across different cohorts. Overall, our results are encouraging for vaccine development since they indicate that even common antibodies can confer specific responses against SARS-CoV-2.

## Discussion

COVID-19 will remain an ongoing threat to public health until an effective SARS-CoV-2 vaccine is available globally. Understanding the human B-cell immune response to SARS-CoV-2 is critical for vaccine development and assessment (Wec et al., 2020a). A repertoire of immune receptor sequences represents a unique snapshot of the history of immune responses in an individual (Boyd et al., 2009; Georgiou et al., 2014; Kreer et al., 2020b; Robins, 2013), and the changes in a repertoire during an infection can signal specific responses to pathogens (Horns et al., 2019; Nourmohammad et al., 2019). Identifying signatures of a functional response to a given pathogen from a pool of mostly unspecific BCRs collected from the blood is challenging—it is a problem of finding a needle in a haystack. Therefore, principled statistical inference approaches are necessary to extract functional signal from such data. Here, we systematically characterize the B-cell repertoire response to SARS-CoV-2 in COVID-19 patients with different disease severity by combining evidence from the overall statistics of repertoires together with dynamics of clonal lineages during infection and the sharing of immune receptors among patients.

At the repertoire level, we showed that the HCDR3 of BCRs in COVID-19 patients are significantly longer than HCDR3 in healthy individuals, and the amino acid composition of this receptor region varies among cohorts of patients with mild, moderate, and severe symptoms. Moreover, we observed large-scale sharing of B-cell receptors among COVID-19 patients, consistent with previous findings in COVID-19 patients (Galson et al., 2020; Nielsen et al., 2020; Schultheiß et al., 2020). Sharing of receptors among individuals can signal common immune responses to a pathogen. However, BCR sharing can also be due to convergent recombination leading to the same receptor sequence or other experimental biases that influence statistics of shared sequences. These statistical nuances can substantially sway conclusions drawn from the sharing analysis and, therefore, should be carefully accounted for. Here, we established a null expectation of BCR sharing due to convergent recombination by inferring a model of receptor generation and migration to the periphery and used this null model to identify sequence outliers. Our analysis identified a subset of rare BCRs shared among COVID-19 patients, which appears to signal convergent responses to SARS-CoV-2.

Bulk B-cell repertoires predominantly contain a mixture of naïve, memory, and plasma-B cells. At the early stages of viral infection, antigen-specific plasma B-cells may develop, which act as antibody factories and confer neutralization against the infecting pathogen (Wrammert et al., 2008). Almost all prior work on immune repertoires has focused on bulk repertoires, which are often easier to sample from and to analyze. Moreover, functional studies, using single-cell sequencing of antigen-sorted B-cell receptors, have often been disconnected from the large-scale analysis of receptor repertoires. Our study synergizes data from bulk and plasma B-cell sequencing with antigen-sorted single-cell B-cell receptors to draw a more complete picture of the human immune response to SARS-CoV-2. Importantly, our joint longitudinal analysis of the bulk and the plasma B-cell repertoires in COVID-19 patients brings insight into the dynamics of antigen-specific B-cells as well as the statistics of receptor sequence features associated with responses to SARS-CoV-2.

In addition to the statistics of repertoires, we observed that the activity of many B-cell lineages (i.e. mRNA production) in COVID-19 patients significantly increases during infection, accompanied by an increase in the binding level ($OD_{450}$) of the patients' plasma to the RBD and NTD of SARS-CoV-2. Dynamics of clonal lineages during an infection provide significant insights into the characteristics of responsive antibodies (Horns et al., 2019; Nourmohammad et al., 2019). By taking advantage of data collected at multiple time points in most patients, we identified expanded lineages shared among patients and found 38 clonal lineages that are candidates for a response specific to SARS-CoV-2 antigens (Fig. 5, Table S5). Importantly, the over-representation of plasma B-cells among these shared expanding lineages signifies their potential role in mounting protective antibody responses against SARS-CoV-2. It should be noted that none of these 38 clonal lineages matched with the verified mAbs. This is in part expected since the verified mAbs that matched the bulk repertoires have relatively high probabilities $P_{\text{post}}$ (Fig. 6C), whereas these 38 lineages are chosen explicitly to be rare.

Our analysis of repertoire dynamics has identified a large-scale expansion of B-cell clonal lineages (5 -15% of lineages) over the course of COVID-19 infections. However, it is hard to imagine that all of these expanding clones that account for a sizeable portion of the repertoire are engaged in responding to SARS-CoV-2 specifically. In contrast, our single-cell analysis identified only about 0.2% of receptors as reactive to RBD and only 0.02% as reactive to NTD epitopes (Fig. S1)—an estimate that is consistent with previous findings (Kreer et al., 2020a). This disparity raises an outstanding question: why do we observe such a large-scale expansion of clonal lineages during an acute immune response?

Identifying antibodies with cross-reactive neutralization abilities against viruses in the SARS family is of significant interest. While cross-neutralization antibodies have been isolated from COVID-19 patients (Brouwer et al., 2020; Liu et al., 2020a; Zhou et al., 2020), it remains unclear how prevalent they are. Interestingly, in nine patients, we see a substantial increase in the binding level ($OD_{450}$) of their plasma to SARS-CoV-1 epitopes during the course of COVID-19 infection. Moreover, in three patients, we identify a BCR identical to the heavy chain of antibody S304 (Pinto et al., 2020), which was previously isolated from a patient who recovered from a SARS-CoV-1 infection. This antibody was shown to be moderately cross-reactive to both SARS-CoV-1 and SARS-CoV-2, and our results further indicate a possibility for such cross-reactive antibodies to emerge naturally in response to SARS-CoV-2 (Brouwer et al., 2020; Lv et al., 2020; Rogers et al., 2020). Taken together, our findings provide substantial insight and strong implications for devising vaccines and therapies with a broad applicability against SARS-CoV-2.

# Materials and Methods

## ▪ Data and code availability

BCR repertoire data and single-cell data can be accessed through:
https://www.ncbi.nlm.nih.gov/bioproject/PRJNA645245
https://www.ncbi.nlm.nih.gov/bioproject/PRJNA679920

All codes for data processing and statistical analysis can be found at: https://github.com/StatPhysBio/covid-BCR

## ▪ Experimental Procedures

**Cell lines.** Sf9 cells (*Spodoptera frugiperda* ovarian cells, female, ATCC catalogue no. CRL-1711) and High Five cells (*Trichoplusia ni* ovarian cells, female; Thermo Fisher Scientific, Waltham, United States (US), catalogue number: B85502) were maintained in HyClone (GE Health Care, Chicago, US) insect cell culture medium.

**Sample collection and PBMC isolation.** Specimens of heparinized blood were collected from the RT-PCR-confirmed COVID-19 patients at the Infectious Disease Centre of the Princess Margaret Hospital, Hong Kong. The study was approved by the institutional review board of the Hong Kong West Cluster of the Hospital Authority of Hong Kong (approval number: UW20-169). All study procedures were performed after informed consent was obtained. Day 1 of clinical onset was defined as the first day of the appearance of clinical symptoms. The severity of the COVID-19 cases was classified based on the adaptation of the Sixth Revised Trial Version of the Novel Coronavirus Pneumonia Diagnosis and Treatment Guidance. The severity of the patients was categorized as follows: Mild - no sign of pneumonia on imaging, mild clinical symptoms; Moderate - fever, respiratory symptoms and radiological evidence of pneumonia; Severe - dyspnea, respiratory frequency >30/min, blood oxygen saturation 93%, partial pressure of arterial oxygen to fraction of inspired oxygen ratio <300, and/or lung infiltrates >50% within 24 to 48 hours; Critical - respiratory failure, septic shock, and/or multiple organ dysfunction or failure or death.

The blood samples were first centrifuged at 3000 xg for 10 minutes at room temperature for plasma collection. The remaining blood was diluted with equal volume of PBS buffer, transferred onto the Ficoll-Paque Plus medium (GE Healthcare), and centrifuged at 400 xg for 20 minutes. Peripheral Blood Mononuclear Cells (PBMC) samples were then collected and washed with cold RPMI-1640 medium for three times. The isolated PBMC samples were finally stored at cell freezing solution (10% DMSO + 90% FBS) and kept in -80$^{o}$C until used.

**RNA extraction and reverse transcription.** Total RNA was extracted from $5 \times 10^5$ PBMC using the RNeasy Mini isolation kit (Qiagen) according to the manufacturer's protocol. Reverse transcription of the RNA samples was performed using the Proto- Script® II Reverse Transcriptase kit (New England Biolabs, NEB) with random hexamer primers according to the manufacturer's protocol. The thermal cycling conditions were designed as follows: 25$^{o}$C for 5 minutes, 42$^{o}$C for 60 minutes, and 80$^{o}$C for 5 minutes. The resulting cDNA samples were stored in 80$^{o}$C freezer before PCR was performed.

**Amplification of B cell repertoire from the samples by PCR.** The cDNA samples were used as a template to amplify the antibody IgG heavy chain gene with six FR1-specific forward primers and one constant region-specific reversed primer using the Phusion® High-Fidelity DNA Polymerase. The primer sequences were the same as previously described (Wu et al., 2015); primer sequences are listed in Table S2. The thermal cycling conditions were set as follows: 98°C for 30 seconds; 30 cycles of 98°C for 10 seconds, 58°C for 15 seconds, and 72°C for 30 seconds; and 72°C for 10 minutes. Then 10 ng of the PCR product was used as a template for the next round of gene amplification with sample-specific barcode primers. The thermal cycling conditions were set as follow: 98°C for 3 min; 30 cycles of 98°C for 10 seconds, 58°C for 15 seconds, and 72°C for 15 seconds; and a final extension at 72°C for 10 min using Phusion® High-Fidelity DNA Polymerase. The PCR product was purified by QIAquick Gel Extraction Kit (Qiagen), and quantified by NanoDrop Spectrophotometers (Thermofisher).

**Protein expression and purification.** The receptor-binding domain (RBD, residues 319–541) and N-terminal domain (NTD, residues 14 to 305) of the SARS-CoV-2 spike protein (GenBank: QHD43416.1) as well as the RBD (residues 306-527) and NTD (residues 14-292) of SARS-CoV-1 spike protein (GenBank: ABF65836.1) were cloned into a customized pFastBac vector (Lv et al., 2020; Wec et al., 2020b). The RBD and NTD constructs were fused with an N-terminal gp67 signal peptide and a C-terminal $His_6$ tag. Recombinant bacmid DNA was generated using the Bac-to-Bac system (Life Technologies, Thermo Fisher Scientific). Baculovirus was generated by transfecting purified bacmid DNA into Sf9 cells using FuGENE HD (Promega, Madison, US) and subsequently used to infect suspension cultures of High Five cells (Life Technologies) at a multiplicity of infection (moi) of 5 to 10. Infected High Five cells were incubated at 28 °C with shaking at 110 rpm for 72 h for protein expression. The supernatant was then concentrated using a Centramate cassette (10 kDa molecular weight cutoff for RBD, Pall Corporation, New York, USA). RBD and NTD proteins were purified by Ni-NTA Superflow (Qiagen, Hilden, Germany), followed by size exclusion chromatography and buffer exchange to phosphate-buffered saline (PBS).

**$CD38^+$ plasma B-cell enrichment**. $CD38^+$ plasma B-cells were isolated from the PBMC samples by performing two subsequent magnetic separation steps according to the manufacturer's protocol (Plasma Cell Isolation Kit II, human, Miltenyi Biotec). Briefly, non-plasma B-cells are labeled with magnetic beads combined with cocktail antibodies and separated using the MACS column. Then, $CD38^+$ plasma B-cell are directly labeled with CD38 MicroBeads and isolated from the pre-enriched B cell pool. Purified $CD38^+$ plasma B-cell were eluted and washed in PBS containing 2% (v/v) fetal bovine serum (FBS) and kept for the following RNA isolation step. In order to test the purity of the $CD38^+$ plasma B cells, we also added staining antibodies and 10 μl of Anti-human CD19-BV510 (BioLegend) and CD38-PE-Cy7 (BioLegend) and incubated them for 15 minutes in the dark in the refrigerator (2-8°C). Cells were finally fixed with 4% PFA for 20 minutes on ice. The stained samples were acquired by flow cytometry on a FACS Attune (Invitrogen) and analyzed with FlowJo software (Fig. S1).

**RBD and NTD protein specific binding B cell enrichment.** B-cells were enriched from the PBMC samples according to the manufacture's protocol (B Cell Isolation Kit II, human, Miltenyi Biotec). Briefly, non-B-cells are labeled with a cocktail of biotin-conjugated antibodies and separated by the MACS column. Purified B-cells were eluted and kept in the PBS buffer with 2% (v/v) FBS. The enriched B cells were then incubated with 2 μg Biotin-RBD or NTD protein for 30 min at 4°C. After incubation, Anti-Biotin MicroBeads were added and incubated for 30 min. RBD and NTD specific bead binding B cells were washed and eluted in PBS and stored on ice until use. In order to test the purity of the RBD- or NTD-specific B cells, we also added staining antibodies, 10 μl of Anti-human CD19-BV510 (BioLegend), and 2 μg of SARS-CoV-2 RBD-PE or NTD-PE and incubated them for one hour in the dark in the refrigerator (2-8°C). Cells were finally fixed with 4% PFA for 20 minutes on ice. The stained samples were acquired by flow cytometry on a FACS Attune (Invitrogen) and analyzed with FlowJo software (Fig. S1).

**Single B cell 5' mRNA and VDJ sequencing.** After RBD or NTD specific B-cells enrichment, cells were counted by using 0.4% (w/v) trypan blue stain solution in the microscope and directly loaded on the 10X Chromium™ Single Cell A Chip. Then single B cell lysis and RNA first-strand synthesis were carried out following the 10X Chromium™ Single Cell 5′ Library & Gel Bead Kit protocol. The RNA sample were used for the next step B cell VDJ library construction following the Chromium™ Single Cell V(D)J Enrichment Kits protocol. VDJ library sequencing was performed on a NovaSeq PE150 and the sequencing data were processed by Cell Ranger.

**ELISA.** A 96-well enzyme-linked immunosorbent assay (ELISA) plate (Nunc MaxiSorp, Thermo Fisher Scientific) was first coated overnight with 100 ng per well of purified recombinant protein in PBS buffer. The plates were then blocked with 100 µl of Chonblock blocking/sample dilution ELISA buffer (Chondrex Inc, Redmon, US) and incubated at room temperature for 1 h. Each human plasma sample was diluted to 1:100 in Chonblock blocking/sample dilution ELISA buffer. Each sample was then added into the ELISA plates for a two-hour incubation at 37°C. After extensive washing with PBS containing 0.1% Tween 20, each well in the plate was further incubated with the anti-human IgG secondary antibody (1:5000, Thermo Fisher Scientific) for 1 hour at 37°C. The ELISA plates were then washed five times with PBS containing 0.1% Tween 20. Subsequently, 100 µL of HRP substrate (Ncm TMB One; New Cell and Molecular Biotech Co. Ltd, Suzhou, China) was added into each well. After 15 min of incubation, the reaction was stopped by adding 50 µL of 2 M $H_2SO_4$ solution and analyzed on a Sunrise (Tecan, Männedorf, Switzerland) absorbance microplate reader at 450 nm wavelength.

## ▪ Statistical Inference and Methods

**BCR preprocessing.** We used a similar procedure for processing of the bulk and the plasma B-cell receptor repertoires. For initial processing of the raw reads, we used pRESTO (version 0.5.13) (Vander Heiden et al., 2014) to assemble paired-end reads, remove sequences with a mean quality score less than 30, mask primer subsequences, and collapse duplicate sequences into unique sequences. The small fraction of paired-end reads that overlapped were assumed to be anomalous and were discarded from the analysis. Additionally, after preprocessing with pRESTO, we discarded unique reads that contained ambiguous calls (N's) in their receptor sequence.

**BCR error correction.** We performed two rounds of error correction on sequences that passed the quality control check. In the first round, we clustered singletons and other low-frequency sequences into larger sequences if they were similar in sequence. The intent of this round was to correct for sequencing errors (e.g. from reverse transcription of mRNA to cDNA) that caused large abundance clones to be split into many similar sequences. We used two parameters: $\Delta_r = 1.0$, the marginal Hamming distance tolerance per decade in log-ratio abundance (each $\log_{10}$ unit allowing $\Delta_r$ additional sequence differences), and $\Delta_a = 1.0$, the marginal abundance tolerance of clusterable sequences per decade in log-ratio abundance (each $\log_{10}$ unit allowing abundance $\Delta_a$ higher as clusterable). For example, a sequence with abundance $a_1$ and a Hamming distance $d$ away from a higher abundance sequence with abundance $a_2$ was absorbed into the latter only if $d \le \Delta_r \log_{10}\frac{a_2}{a_1}$ and $a_1 \le \Delta_a \log_{10}\frac{a_2}{a_1}$. We used the output of this first round as input for the second round of error correction, in which we more aggressively target correction of reverse transcriptase errors. In the second round, we used two different parameters to assess sequence similarity: $d_{\text{thresh}} = 2.0$, the Hamming distance between sequences, and $a_{\text{thresh}} = 1.0$, the ratio of sequence abundances. A sequence with abundance $a_1$ and a Hamming distance $d$ away from a sequence of larger abundance $a_2$ was absorbed into the latter only if $d \le d_{\text{thresh}}$ and the ratio of the sequence abundances was greater than $a_{\text{thresh}}$, i.e. $\frac{a_2}{a_1} \ge a_{\text{thresh}}$. This round of error correction allows much larger abundance sequences to potentially be clustered than is possible in the first round. For both of the above steps, we performed clustering greedily and approximately by operating on sequences sorted by descending abundance, assigning the counts of the lower abundance sequence to the higher abundance one iteratively.

After error correction, the sequences still contained a large number of singletons, i.e. sequences with no duplicates (Tables S1, S4). We discarded these singletons from all analyses that relied on statistics of unique sequences (i.e., the results presented in Figs. S2A-C and S3E-G).

**BCR annotation.** For each individual, error-corrected sequences from all timepoints and replicates were pooled and annotated by abstar (version 0.3.5) (Briney and Burton, 2018). We processed the output of abstar, which included the estimated IGHV gene/allele, IGHJ gene/allele, location of the HCDR3 region, and an inferred naïve sequence (germline before hypermutation). Sequences which had indels outside of the HCDR3 were discarded. We partitioned the sequences into two sets: productive BCRs, which were in-frame and had no stop codons, and unproductive BCRs, which were out-of-frame.

**Unproductive BCRs.** Due to a larger sequencing depth in healthy individuals, we were able to reconstruct relatively large unproductive BCR lineages. Unproductive sequences are BCRs that are generated but, due to a frameshift or insertion of stop codons, are never expressed. These BCRs reside with productive (functional) BCRs in a nucleus and undergo hypermutation during B-cell replication and, therefore, provide a suitable null expectation for generation of BCRs in immune repertoires.

**Clonal lineage reconstruction.** To identify BCR clonal lineages, we first grouped sequences by their assigned IGHV gene, IGHJ gene, and HCDR3 length and then used single-linkage clustering with a threshold of 85% Hamming distance. A similar threshold has been suggested previously by (Gupta et al., 2017) to identify BCR lineages. Defining size as the sum of the number of unique sequences per time point within a lineage, clusters of size smaller than three were discarded from most analyses. They were retained only for training IGoR and SONIA models and were not discarded in the sharing analysis only if the progenitor of that small cluster was also a progenitor of a cluster of size at least three in another patient. For each cluster, there may have been multiple inferred naïve sequences, as this was an uncertain estimate. Therefore, the most common naïve sequence was chosen to be the naïve progenitor of the lineage. When the most common naïve sequence of a productive lineage contained a stop codon, the progenitor of the lineage was chosen iteratively by examining the next most common naïve sequence until it did not contain any stop codons. If all inferred naïve sequences in a productive lineage had a stop codon, that lineage was discarded from the analysis. Tables S1 and S4 show the statistics of constructed clonal lineages in each individual for the bulk repertoire and combined bulk+plasma B-cell repertoire, respectively.

**Mapping of single-cell data onto reconstructed clonal lineages:** Like the repertoire datasets, the single-cell sequences were annotated by abstar (Briney and Burton, 2018). For each receptor acquired by single-cell sequencing, we identified a subset of reconstructed clonal lineages from the bulk repertoire which had identical HCDR3 length as the sequence and which also had an IGHV gene which was 90% similar to that of the single-cell receptor. This flexibility in V-gene choice would identify functionally homologous receptors and associate a receptor to a lineage with a sequence divergence in the V-segment, compatible with the expectation under somatic hypermutations (Lee et al., 2017). A single-cell sequence was matched to a reconstructed clonal lineage from this subset if its HCDR3 could be clustered with other members of the lineages, using single-linkage clustering with a similarity threshold of 85% Hamming distance (similar to the criteria for lineage reconstruction for bulk repertoires).

**Inference of generation probability and selection for BCRs.** We used IGoR (version 1.4) (Marcou et al., 2018) to obtain a model of receptor generation. This model characterized the probability of generation $P_{\mathrm{gen}}(\sigma)$ of a receptor dependent on the features of the receptor, including the IGHV, IGHD, and IGHJ genes and the deletion and insertion profiles at the VD and DJ junctions. To characterize the parameters of this model, we trained IGoR on the progenitors of unproductive lineages, regardless of size, pooled from the bulk repertoire of all individuals, restricted to progenitors whose HCDR3 began with a cysteine and ended with a tryptophan. For consistency with our receptor annotations based on abstar, we used abstar's genomic templates and the HCDR3 anchors of abstar's reference genome as inputs for IGoR's genomic templates and HCDR3 anchors. $P_{\mathrm{gen}}(\sigma)$ distributions of the healthy and COVID-19 cohorts in this study are shown in Fig. S4A.

We used SONIA (version 0.45) (Sethna et al., 2020) to infer a selection model for progenitors of productive clonal lineages. The SONIA model evaluated selection factors $q$ to characterize the deviation in the probability $P_{\mathrm{post}}(\sigma)$ to observe a functional sequence in the periphery from the null expectation based on the generation probability $P_{\mathrm{gen}}(\sigma)$: $P_{\mathrm{post}}(\sigma) = \frac{1}{Z} P_{\mathrm{gen}}(\sigma) e^{\sum_{f:\mathrm{features}} q_f(\sigma)}$, where $Z$ is the normalization factor and $q_f(\sigma)$ are selection factors dependent on the sequence features $f$. These sequence features include IGHV-gene and IGHJ-gene usages and HCDR3 length and amino acid composition (Sethna et al., 2020).

In our analysis, we used the SONIA left-right model with independent IGHV- and IGHJ-gene usages (Sethna et al., 2020). We used the output from IGoR as the receptor generation model for SONIA. We trained four cohort-specific SONIA models on progenitors of productive lineages, regardless of size, pooled from the bulk repertoire of all individuals within a cohort, restricted to progenitors whose HCDR3 began with a cysteine and ended with a tryptophan. 150 epochs, $L_2$ regularization with strength 0.001, and 500,000 generated sequences were used to train each SONIA model. Fig. 3 shows the distributions for the probabilities of observing productive receptors sampled from each cohort $P_{\text{post}}(\sigma)$ and the correlation of feature-specific selection factors $q_f$ among cohorts. A SONIA model was also trained on all the productive lineage progenitors in the GRP dataset (Briney et al., 2019) and used 5,000,000 generated sequences, keeping the other parameters unchanged. We refrain from comparing directly $P_{\text{post}}(\sigma)$ associated with GRP BCRs to BCRs in this study due to experimental differences.

It should be noted that the (pre-selection) generation model $P_{\text{gen}}(\sigma)$ inferred by IGoR (Marcou et al., 2018) is robust to sequence errors due to experimental errors or hypermutations in the IgG repertoires. However, hypermutations in BCRs could introduce errors in inference of selection models and estimation of receptor probabilities by SONIA (Sethna et al., 2020). Therefore, we have restricted our selection analyses to only the inferred progenitors of clonal lineages. Although the inferred progenitors of lineages can still deviate from the *true* (likely IgM naïve) progenitors, the selection models inferred from *ensembles* of inferred progenitors in IgG repertoires seem to be comparable to the models inferred from the IgM repertoires (Ruiz Ortega et al., 2021). The resulting selection models, trained on either true or inferred progenitors, reflect preferences for sequence features of unmutated receptors, including IGHV- and IGHJ- genes and HCDR3 length and composition, but they do not account for the hypermutation preferences that may distinguish one cohort from another.

**Characterizing the robustness of selection inference.** To test the sensitivity of the inferred selection models on the size of the training sets, we down-sampled the receptor data of each COVID-19 cohort to a size comparable to the smallest cohort, i.e., the healthy repertoire sequenced in this study. This down-sampling resulted in two independent training datasets for the mild COVID-19 cohort, 13 independent training datasets for the moderate COVID-19 cohort, and three independent training datasets for the severe COVID-19 cohort. Though this down-sampling resulted in over 400 independent training datasets for the GRP, we elected to use only 15. We then inferred a separate selection model with SONIA for each of these training datasets and used each model to evaluate the receptor log-probabilities $\log_{10} P_{\text{post}}(\sigma)$ for a set of 500,000 generated receptors. The evaluated probabilities are strongly correlated between models inferred from the down-sampled data in each cohort, with a Pearson correlation of $r > 0.99$ and p-value = 0 (Fig. S4C-F).

We used a similar approach to compare the selection model inferred from the healthy repertoires sequenced in this study and the GRP study (Briney et al., 2019). Fig. S4B shows that, using the model inferred with our healthy repertoire and 30 down-sampled independently inferred selection models using the GRP dataset, the evaluated log-probabilities $\log_{10} P_{\text{post}}(\sigma)$ based on these two datasets are strongly correlated, with a Pearson correlation of $r > 0.99$ and p-value = 0 (Fig. S4B).

**Characterizing repertoire diversity.** We quantified the diversity of each cohort by evaluating the entropy of receptor sequences in each cohort. Entropy can be influenced by the size of the training dataset for the selection models. To produce reliable estimates of repertoires' diversities (and entropies), we used the procedure described above to learn independent selection models for subsampled repertoires in each cohort. We then used the inferred IGoR and SONIA models to generate 500,000 synthetic receptors based on each of the subsampled, cohort-specific models. We evaluated cohort entropies $H$ as the expected log-probabilities to observe a functional sequence in the respective cohort: $H = -\sum_{\sigma} P_{\text{post}}(\sigma) \log P_{\text{post}}(\sigma)$; the estimates based on the generated receptors are reported in the main text. The error bars reported for these entropy estimates are due to variations across the inferred models in each cohort.

For comparison, we also evaluated the entropy estimated on the repertoire data in each cohort, which showed a similar pattern to the estimates from the generated cohorts (in the main text). Specifically, the entropy of BCR repertoires estimated from the data follows: $39.8 \pm 0.3$ bits in healthy individuals, $41.9 \pm 0.7$ bits for patients in the mild cohort, $42.7 \pm 0.3$ bits for patients in the moderate cohort, and $42.9 \pm 0.5$ for patients in the severe cohort. The error bars indicate the standard error due to differences among individuals within a cohort.

**Comparing selection between repertoires of cohorts.** Selection models enable us to characterize the sequence features of immune repertoires that differ between cohorts. We evaluated the Jensen-Shannon divergence $D_{JS}(r, r')$ between the distribution of repertoires $r$ and $r'$, $P^{r}_{\text{post}}$ and $P^{r'}_{\text{post}}$, defined as

$$
\begin{aligned}
D_{JS}(r,r') &= \frac{1}{2}\sum_{\sigma:\,\text{sequences}} P^{r}_{\text{post}}(\sigma)\log\frac{P^{r}_{\text{post}}(\sigma)}{\left(P^{r}_{\text{post}}(\sigma)+P^{r'}_{\text{post}}(\sigma)\right)/2} + \frac{1}{2}\sum_{\sigma:\,\text{sequences}} P^{r'}_{\text{post}}(\sigma)\log\frac{P^{r'}_{\text{post}}(\sigma)}{\left(P^{r}_{\text{post}}(\sigma)+P^{r'}_{\text{post}}(\sigma)\right)/2} \\
&= \frac{1}{2}\sum_{\sigma:\,\text{sequences}} P^{r}_{\text{post}}(\sigma)\log\frac{2\,Q^{r}(\sigma)}{Q^{r}(\sigma)+Q^{r'}(\sigma)} + \frac{1}{2}\sum_{\sigma:\,\text{sequences}} P^{r'}_{\text{post}}(\sigma)\log\frac{2\,Q^{r'}(\sigma)}{Q^{r}(\sigma)+Q^{r'}(\sigma)}
\end{aligned}
$$

where we used the relationship between a receptor's generation probability $P_{\text{gen}}(\sigma)$ and its probability after selection $P^{r}_{\text{post}}(\sigma)$, using the inferred selection factor $Q^{r}(\sigma) = \frac{1}{Z}\, e^{\sum_{f:\text{features}} q^{r}_{f}(\sigma)}$ in repertoire $r$: $P^{r}_{\text{post}}(\sigma) = P_{\text{gen}}(\sigma)\, Q^{r}(\sigma)$. The Jensen-Shannon divergence $D_{JS}(r,r')$ is a symmetric measure of distance between two repertoires, which we can calculate using their relative selection factors (Isacchini et al., 2021). Fig. 3 shows the expected partial Jensen-Shannon divergences evaluated over five independent realizations of 100,000 generated sequences for each partial selection model. The error bars show the variations of these estimates over the five independent realizations in this procedure.

**Clonal lineage expansion.** We studied clonal lineage expansion of BCR repertoires in individuals that showed an increase in the binding level ($OD_{450}$) of their plasma to SARS-CoV-2 (RBD) during infection (Figs. 5A, S5): patients 2, 3, 4, 5, 6, 7, 9, 10, 11, 13, 14. Other individuals showed no increase in IgG binding to SARS-CoV-2 (RBD), either due to already high levels of binding at early time points or to natural variation and noise (Fig. S5). Our expansion test compared two time points. Therefore, for individuals with three time points, we combined data from different time points such that the separated times coincided with larger changes in binding levels ($OD_{450}$). Specifically, we combined the last two time points for patients 2 and 7 and the first two time points for patient 9. In addition, we combined replicates at the same time point and filtered out small lineages with size less than three, where size was defined as the sum of the amount of unique sequences per time-point within a lineage.

To test for expansion, we compared lineage abundances (i.e., total number of reads in a lineage) between early and late time points. Many lineages appeared only in one time point due to the sparse sampling of clonal lineages and the cells that generate them (Fig. S8). Therefore, we tested for expansion only for lineages that had nonzero abundances at both time points.

Our expansion test relied on comparing the relative abundance of a given lineage with other lineages. However, due to primer-specific amplification biases, abundances were not comparable between reads amplified with different primers. Therefore, in our analysis we only compare a lineage with all other lineages that were amplified with the same primer.

We applied a hypergeometric test (Fisher's exact test) to characterize significance of abundance fold change for a focal lineage. A similar method was used to study clonal expansion in TCRs (DeWitt et al., 2015). For each focal clonal lineage $i$ (in a given individual), we defined a $2 \times 2$ contingency matrix $C$,

$$\mathrm{C} = \begin{pmatrix} n_i^{\mathrm{early}} & N_{/i}^{\mathrm{early}} \\ n_i^{\mathrm{late}} & N_{/i}^{\mathrm{late}} \end{pmatrix}$$

where $n_i^{\mathrm{early}}$ and $n_i^{\mathrm{late}}$ are the abundances of the focal lineage at the early and late time, and $N_{/i}^{\mathrm{early}}$ and $N_{/i}^{\mathrm{late}}$ are the total abundances of all reads (with the same primer) minus those from lineage $i$ at the early and late times. The ratio $\frac{n_i^{\mathrm{late}} / n_i^{\mathrm{early}}}{N_{/i}^{\mathrm{late}} / N_{/i}^{\mathrm{early}}}$ describes the fold change, or odds ratio, of lineage $i$ relative to the rest of the reads in the same primer group. Based on the contingency matrix $C$, one-sided p-values for Fisher's exact test were calculated using the "fisher.test" function in R version 4.0. Fold change and p-values are shown in Fig. S6G.

To determine a significance threshold for the Fisher's exact test, we examined the replicate data from samples collected from the same time point in each individual because we did not expect any significant expansion among replicates. We performed the expansion test on pairs of replicates (Fig. S6C) and compared the empirical cumulative distributions of the time point and replicate expansion data (Fig. S6E,F) (Storey, 2002; Storey and Tibshirani, 2003). We chose a p-value threshold of $10^{-300}$, where there were 12.3 as many significant expansions as in the replicate data, and therefore the false discovery rate was approximately $1/(1 + 12.3) = 0.075$.

**Significance of BCR sharing among individuals.** The probability that receptor $\sigma$ is shared among a given number of individuals due to convergent recombination can be evaluated based on the probability to observe a receptor in the periphery $P_{\text{post}}(\sigma)$, the size of the cohort $M$, and the size of the repertoire (sequence sample size) $N$. First, we evaluated the probability $\rho(\sigma; N)$ that receptor $\sigma$ with probability $P_{\text{post}}(\sigma)$ appears at least once in a sample of size $N$,

$$\rho(\sigma; N) = 1 - \left(1 - P_{\text{post}}(\sigma)\right)^{N} \simeq 1 - e^{-NP_{\text{post}}}$$

The probability that receptor $\sigma$ is shared among $m$ individuals out of a cohort of $M$ individuals, each with a (comparable) sample size $N$, follows the binomial distribution,

$$P_{\text{share}}(\sigma; m, M, N) = \binom{M}{m} [\rho(\sigma; N)]^{m} [1 - \rho(\sigma; N)]^{M-m}$$

We aimed to identify shared receptors that were outliers such that their probability of sharing is too small to be explained by convergent recombination or other biases in the data. To do so, we identified the receptors with the smallest sharing probabilities $P_{\text{share}}$ and found a threshold of $P_{\text{post}}$ (dashed lines in Fig. 6 and Fig. S11) at the 2% quantile of $P_{\text{share}}$ in the data. Specifically, since $P_{\text{share}}$ is a function of $P_{\text{post}}$ and $m$ (number of individuals sharing), for each $m$ we solved for $P_{\text{post}}$ such that $P_{\text{share}} = c$, and tuned the constant $c$ such that only 2% of the data lay below $P_{\text{share}}$. This was a conservative choice to identify the rare shared outliers in the data.

## Acknowledgments

This work was supported by DFG grant (SFB1310) on Predictability in Evolution (A.N., Z.M., J.O., G.I.), the Max Planck Society through MPRG funding (A.N., Z.M., J.O., G.I.), Department of Physics at the University of Washington (A.N., Z.M.), Royalty Research Fund at the University of Washington (A.N., Z.M.), NIH NIAID F31AI150163 (WSD), Calmette and Yersin scholarship from the Pasteur International Network Association (H.L.), Bill and Melinda Gates Foundation OPP1170236 (I.A.W.), a startup fund at the University of Illinois at Urbana-Champaign (N.C.W.),

US National Institutes of Health (contract no. HHSN272201400006C) (J.S.M.P), National Natural Science Foundation of China (NSFC)/Research Grants Council (RGC) Joint Research Scheme (N_HKU737/18) (C.K.P.M. and J.S.M.P) and the Research Grants Council of the Hong Kong Special Administrative Region, China (Project no. T11-712/19-N) (J.S.M.P). We acknowledge the support of the clinicians who facilitated this study, including Drs Wai Shing Leung, Jacky Man Chun Chan, Thomas Shiu Hong Chik, Chris Yau Chung Choi, John Yu Hong Chan, Daphne Pui-Lin Lau, and Ying Man Ho; the dedicated clinical team at Infectious Diseases Centre, Princess Margaret Hospital, Hospital Authority of Hong Kong; and the patients who kindly consented to participate in this investigation. We also thank the Center for PanorOmic Sciences (CPOS), LKS Faculty of Medicine, and University of Hong Kong for their support on next-generation sequencing and acknowledge the use of the computational infrastructure provided by the Hyak supercomputer system funded by the student technology fund (STF) at the University of Washington.

## Author Contributions

Z.M., H.Lv, J. O, I.A.W., J.S.M.P. N.C.W., A.N., and C.K.P.M. conceived and designed the study. O.T.-Y.T. organized patient recruitment, data collection, and sampling. H.Lv, G.K.Y., W.W.N., and C.K.P.M. prepared the next-generation sequencing libraries and performed the ELISA experiments. M.Y., H.Liu, and N.C.W. and expressed and purified the proteins. Z.M., J.O., W.S.D., G.I., and A.N. analyzed the data and performed the modelling work and statistical inference. Z.M., H.Lv, N.C.W., A.N., and C.K.P.M. wrote the paper. All authors reviewed and edited the paper.

## Competing Interests

The authors declare no competing interests.

# Supplementary Information

## Dynamics of B-cell repertoires and emergence of cross-reactive responses in COVID-19 patients with different disease severity

Montague *et. al.*

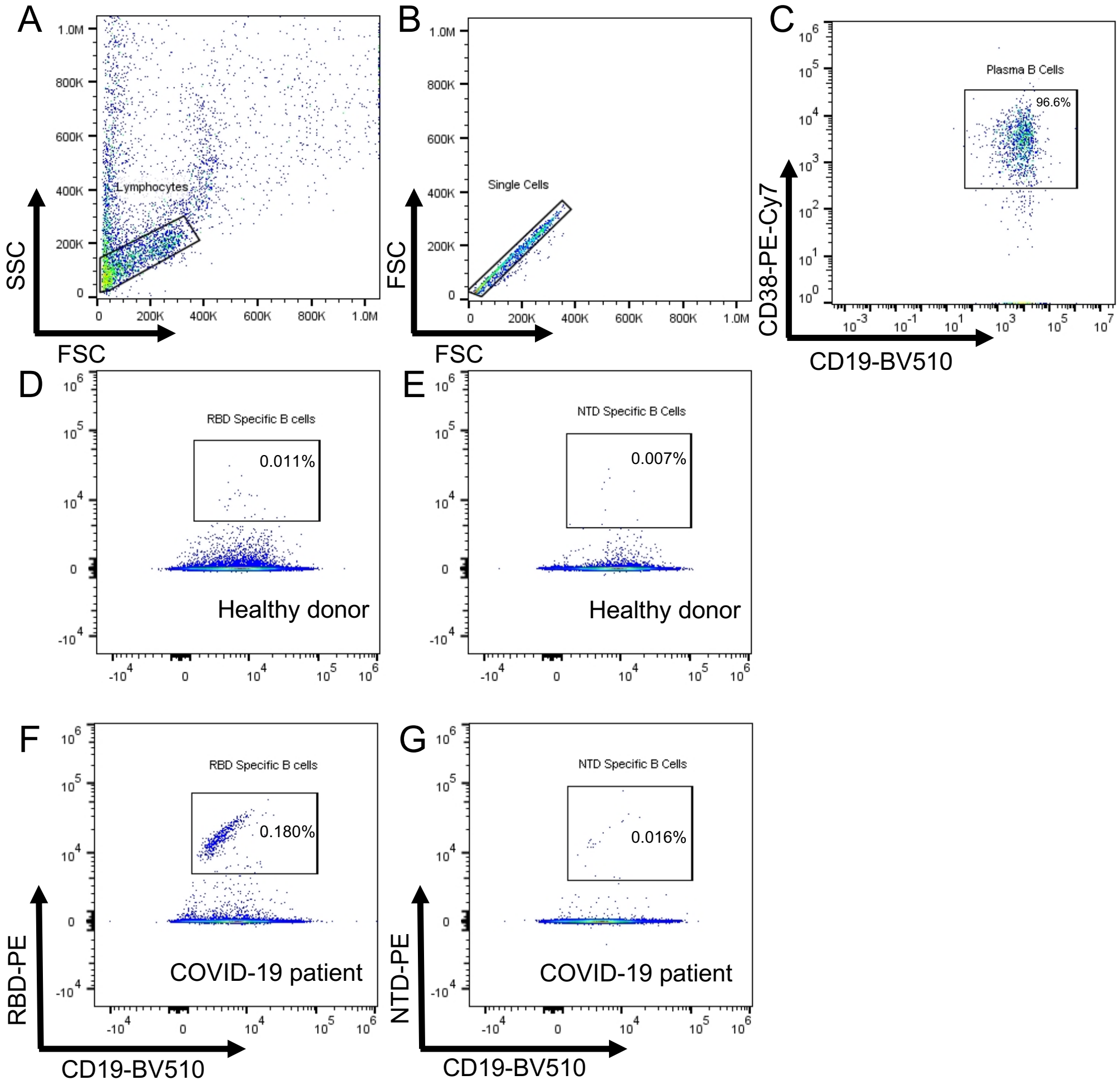

**Figure S1. Gating strategy for the CD38+ Plasma B cell. (A-C)** Flow cytometry stainings of CD38+ plasma B cells from CDVID-19 infected patient using two fluorescent markers, anti-human CD19-BV510 (BioLegend) and CD38-PE-Cy7 (BioLegend) in the same tube. Percentages indicate the proportions of $CD19^{+}CD38^{+}$ plasma B cells within total B cells. **(D-G)** Flow cytometry stainings of RBD- or NTD-specific B cell from healthy donor and CDVID-19 infected patient using anti-human CD19-BV510 (BioLegend) and PE fluorescent dye for RBD protein (D, F) or NTD protein (E, G) in the same staining tube. Percentages indicate the proportions of $CD19^{+}$ and RBD- or NTD-protein double positive specific B cells.

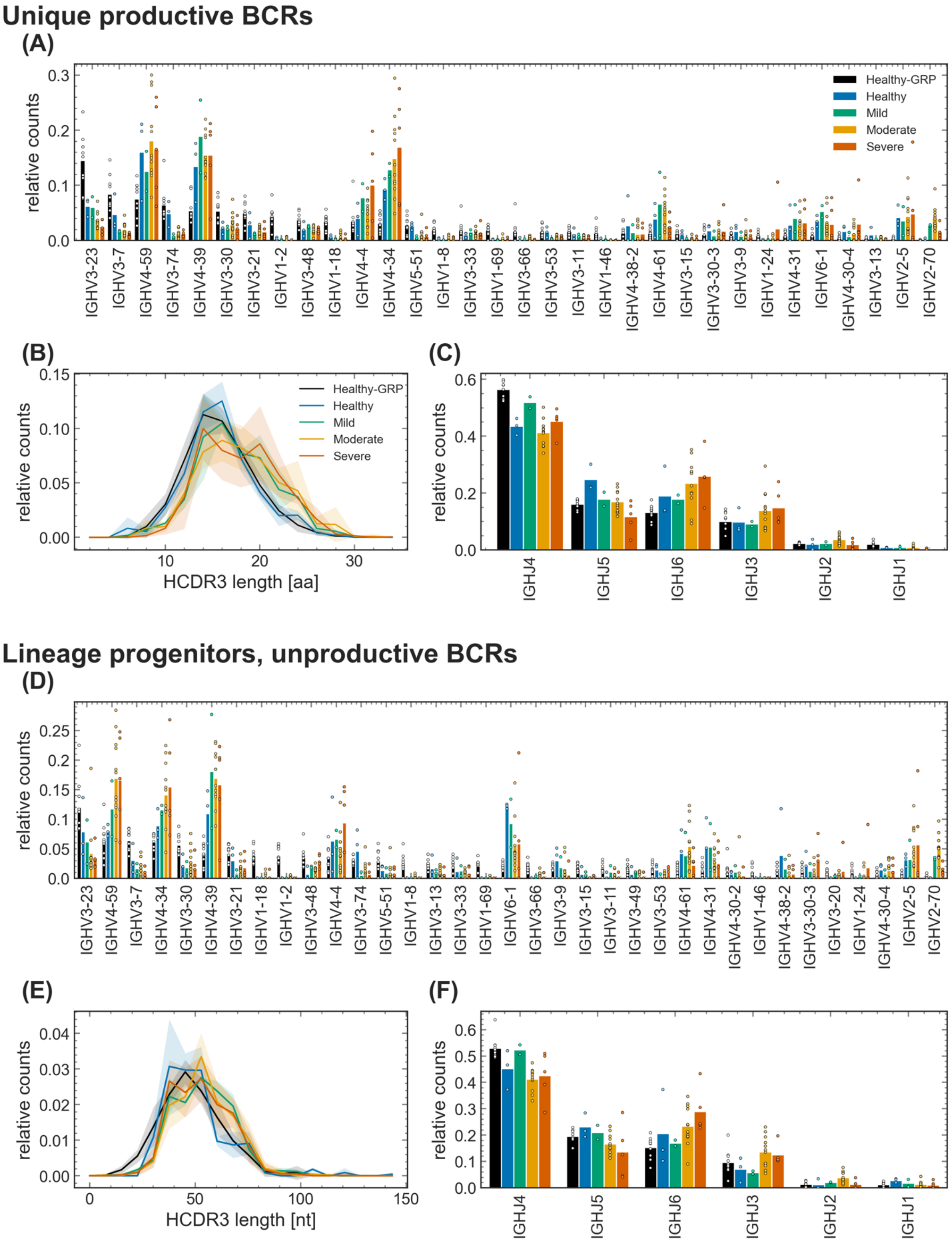

**Figure S2. Bulk repertoire sequence statistics. (A-C)** Similar statistics are shown as in Fig. 3 (A, C-D), but for unique receptors excluding singletons (Methods). Unique BCRs in healthy individuals (our control and the Great Repertoire Project (GRP) by (Briney et al., 2019) show significantly shorter HCDR3s compared to moderate and severe cohorts. ANOVA statistics for mean HCDR3 length between cohorts: Healthy-Mild: $F_{1,3} = 8.7$, p-value = 0.06; Healthy-Moderate: $F_{1,13} = 17.2$, p-value = 0.001; Healthy-Severe: $F_{1,6} = 10.0$, p-value = 0.020; GRP-Mild: $F_{1,10} = 11.3$, p-value = 0.0073; Healthy-GRP: $F_{1,11} = 0.074$, p-value = 0.791; GRP-Moderate: $F_{1,20} = 34.0$, p-value = 0.000011; GRP-Severe: $F_{1,13} = 41.5$, p-value = 0.000022. **(D-F)** Similar statistics are shown as in Fig. 2 (A, C-D), but for unproductive lineage progenitors. The differences in the statistics of HCDR3 length between the unproductive repertoires of healthy individuals and the COVID-19 cohorts are insignificant (ANOVA p-value > 0.01). Colors are consistent across panels.

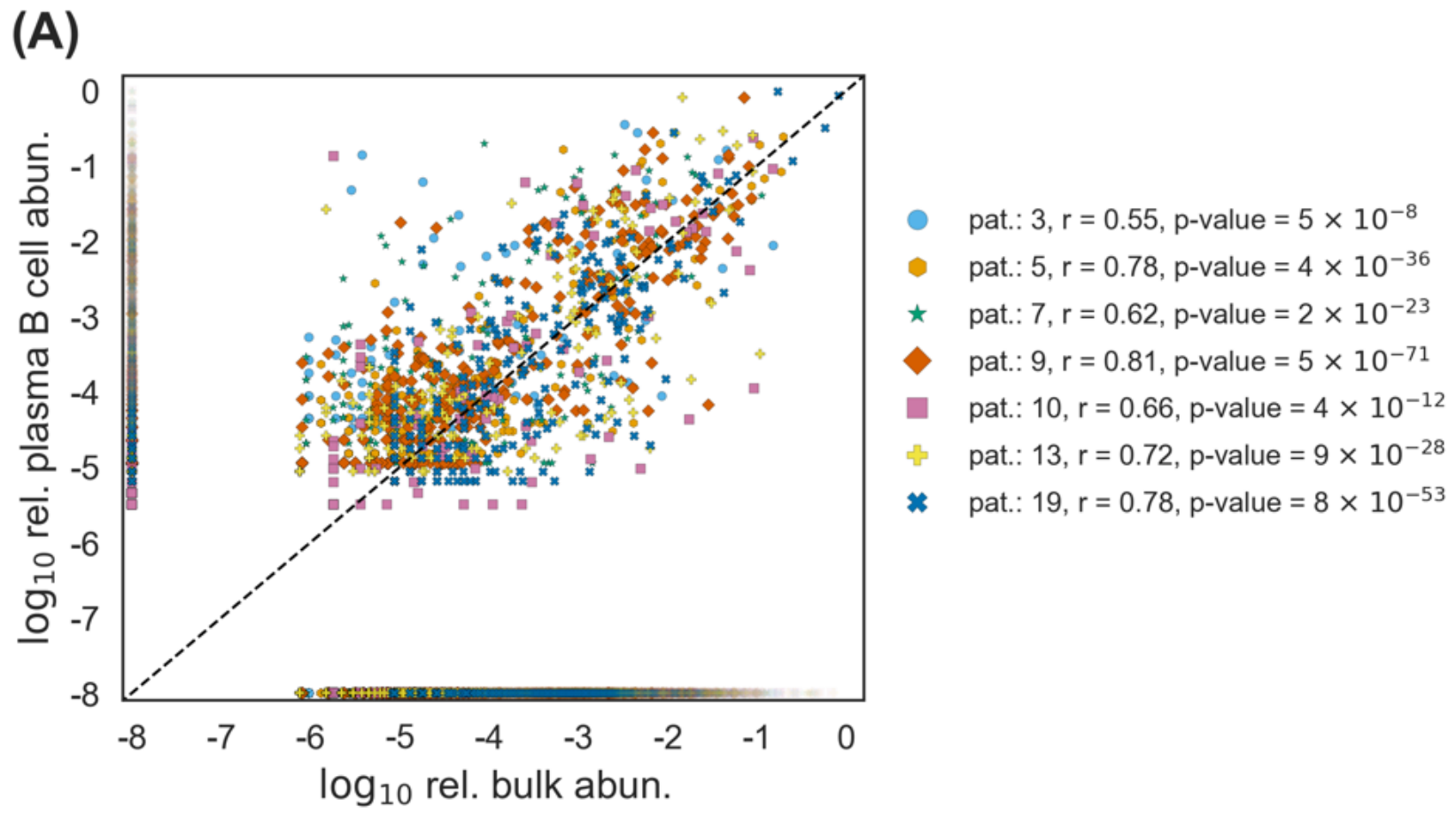

## Lineage progenitors, productive BCRs

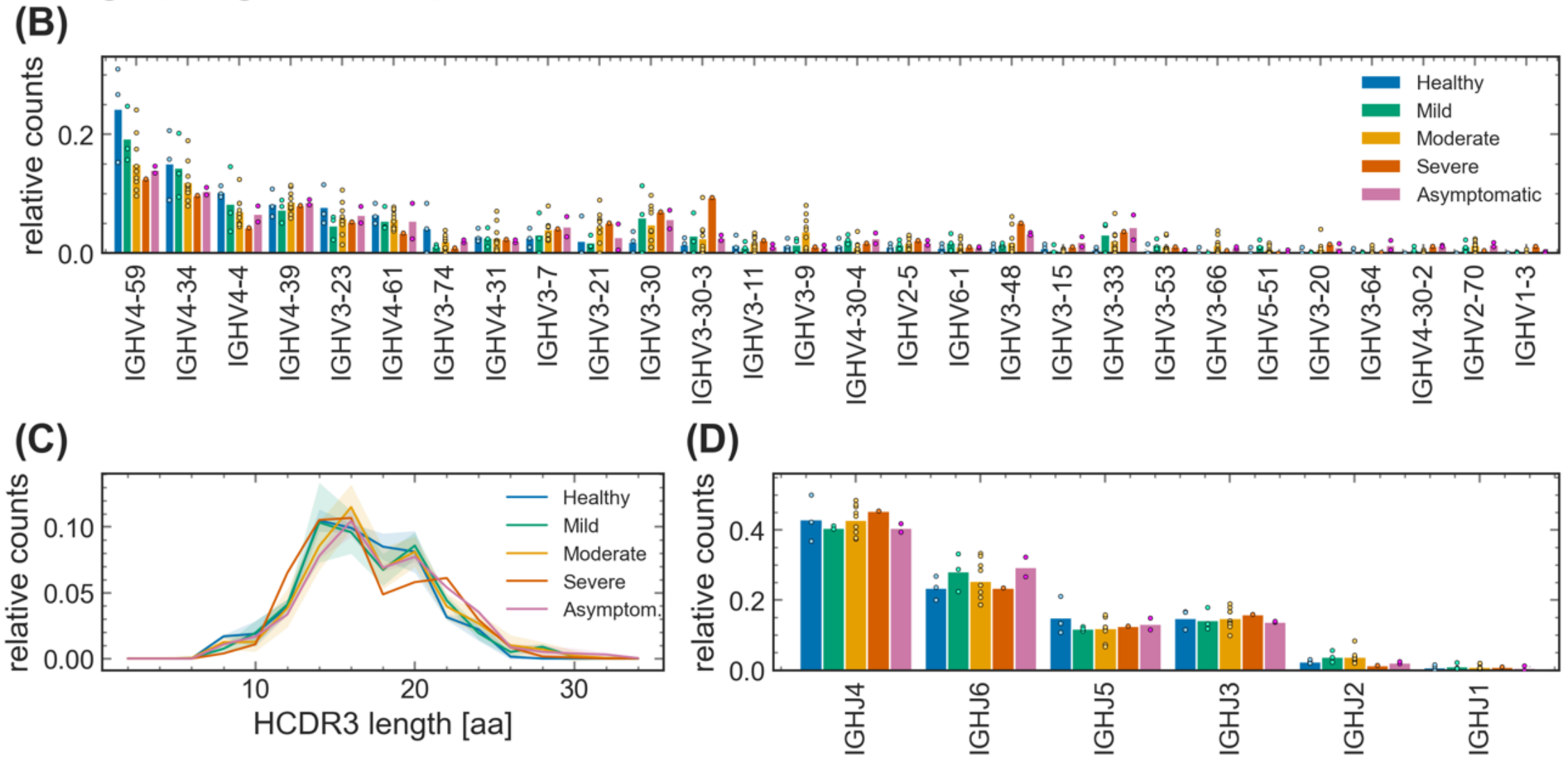

## Unique productive BCRs

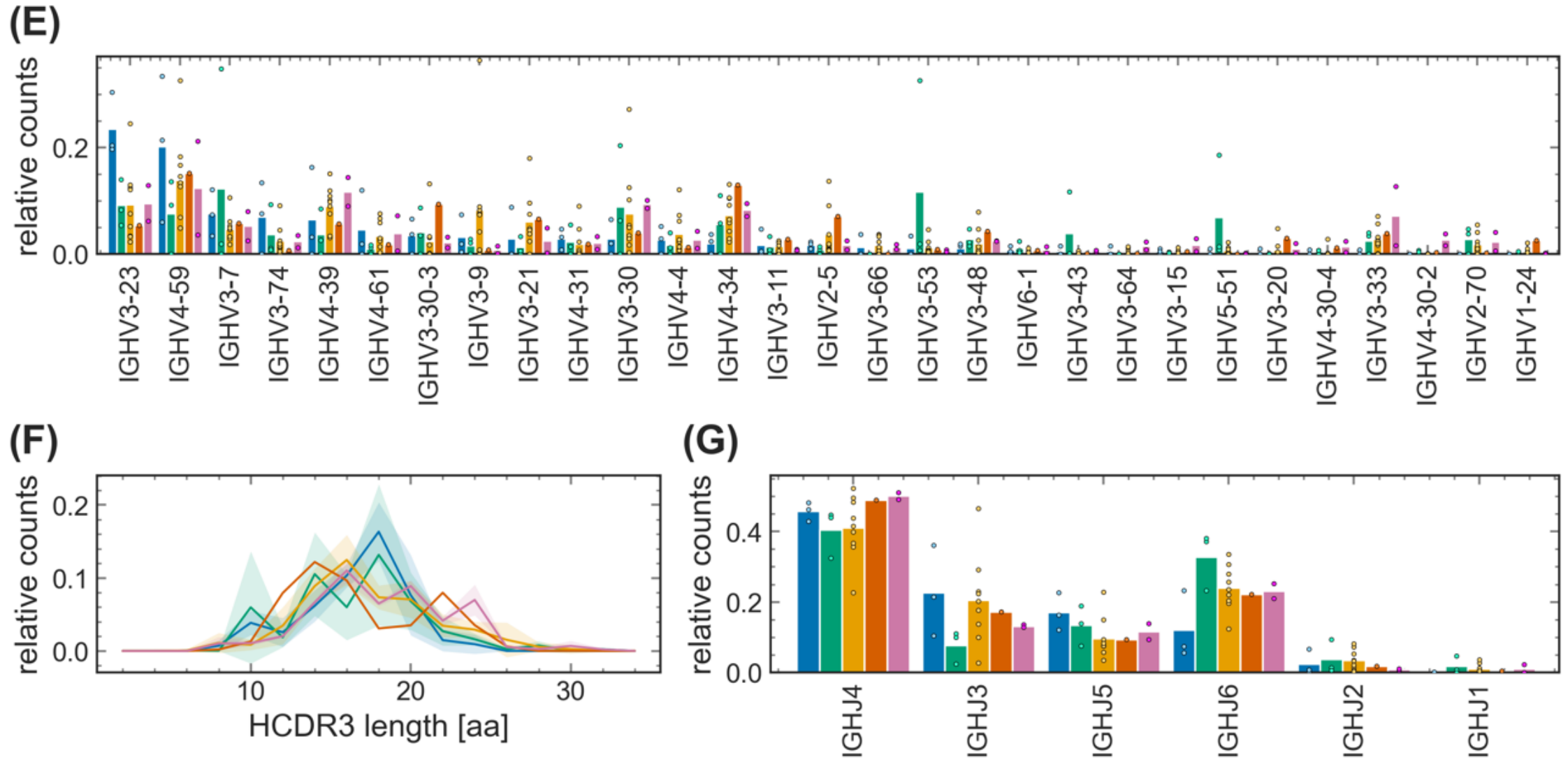

**Figure S3. Sequence features of immune receptors in the plasma B-cell repertoire across cohorts. (A)** Scatter plot shows $\log_{10}$ relative abundance of clonal lineages constructed from the plasma B-cell and bulk repertoire data from all time points and replicates in each patient (colors). To avoid primer-specific amplification biases, the relative abundance is estimated as the total read count of a clonal lineage relative to the total reads in the data associated with a specific primer amplification. Lineages with only bulk reads or only plasma reads are displayed as having $\log_{10}$ relative abundance = 1e-8. Pearson correlations (r) between abundances of lineages which were present in both the bulk and the plasma B-cell repertoires and the corresponding p-values are indicated in the legend for each patient. **(B-D)** Similar statistics are shown as in Fig. 3 (A,C,D), but for progenitors of clonal lineages with minimum size of three, in which at least one BCR is found in the plasma B-cell repertoire data; statistics of these lineages are reported in Tables S1, S2. Smaller read counts in the plasma B-cell data compared to the bulk do not allow for comparative analysis of receptor statistics across cohorts. **(D-F)** Similar statistics are shown as in Fig. S2 (A-C), but for unique receptors harvested from the plasma B-cell repertoires. Statistics of these receptors in each individual is described in Table S2. Smaller read counts in the plasma B-cell data compared to the bulk don’t allow for comparative analysis of receptor statistics across cohorts. Colors are consistent across panels.

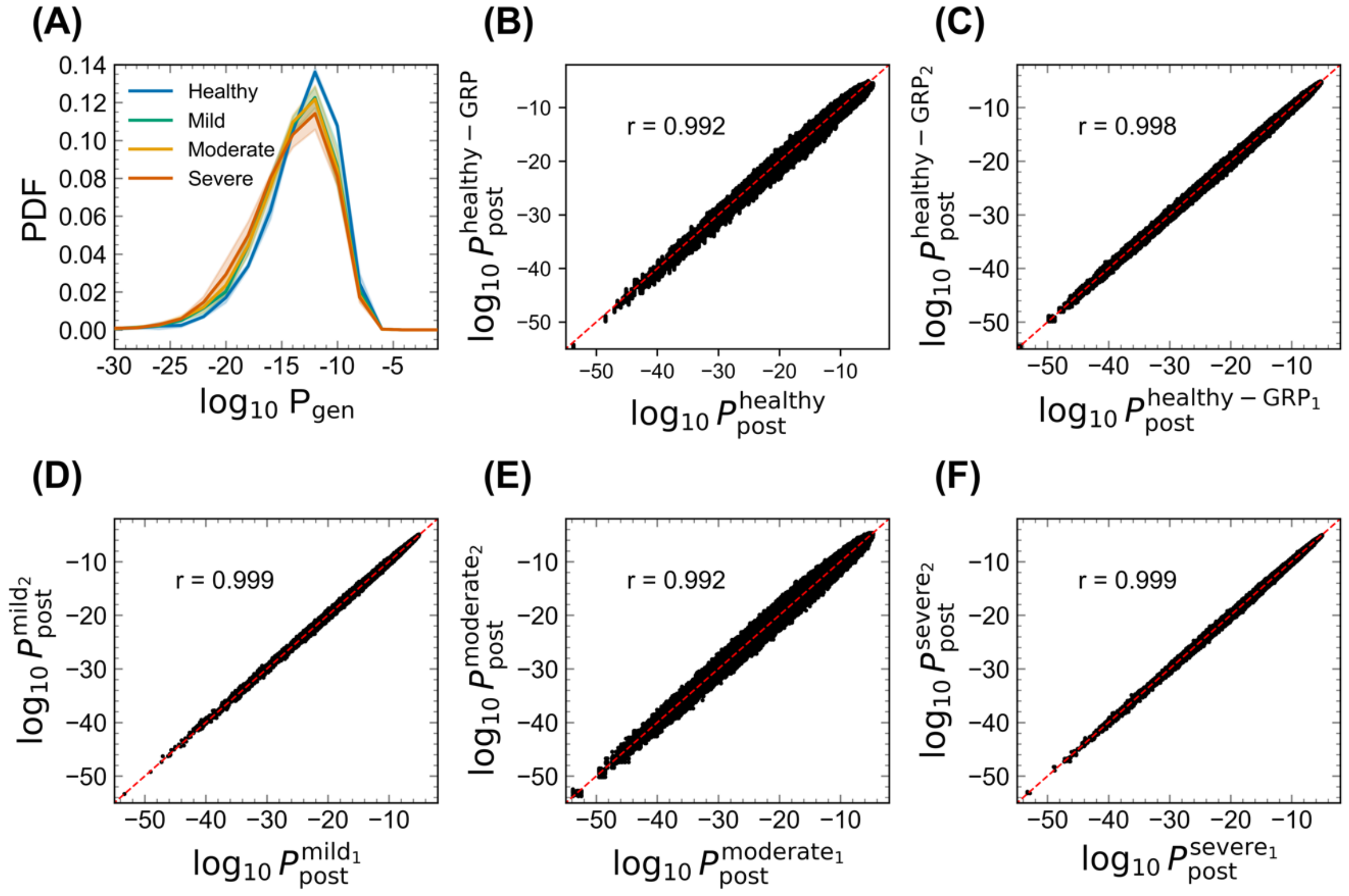

**Figure S4. Robustness of SONIA selection models. (A)** The distribution of the log-generation probability of a sequence $\sigma$ $\log_{10} \mathrm{P_{gen}}(\sigma)$, evaluated using the inferred generation models by the IGoR software (Marcou et al., 2018), is shown as a normalized probability density function (PDF) for inferred naïve progenitors of productive clonal lineages in cohorts of healthy individuals and the mild, moderate, and severe cohorts of COVID-19 patients (colors). Full lines show distributions averaged over individuals in each cohort, and shadings indicate regions containing one standard deviation of variation among individuals within a cohort. **(B)** The scatterplot shows log $P_{\text{post}}$ obtained by evaluating 500,000 generated sequences using the inferred selection (SONIA) models (Sethna et al., 2020) trained on the healthy cohort (x-axis) and 30 SONIA models trained on independent samples of the GRP dataset (Briney et al., 2019) down-sampled to the size of the healthy cohort in this study (7,161 receptors) (y-axis). The scatterplots show all unique pairwise comparisons between SONIA models trained on independent subsets with each cohort for **(C)** GRP (30 models), and COVID-19 patients with **(D)** mild (two models), **(E)** moderate (13 models), and **(F)** sever (three models) symptoms (Methods). The Pearson correlation between for pairwise model comparisons are shown in each panel.

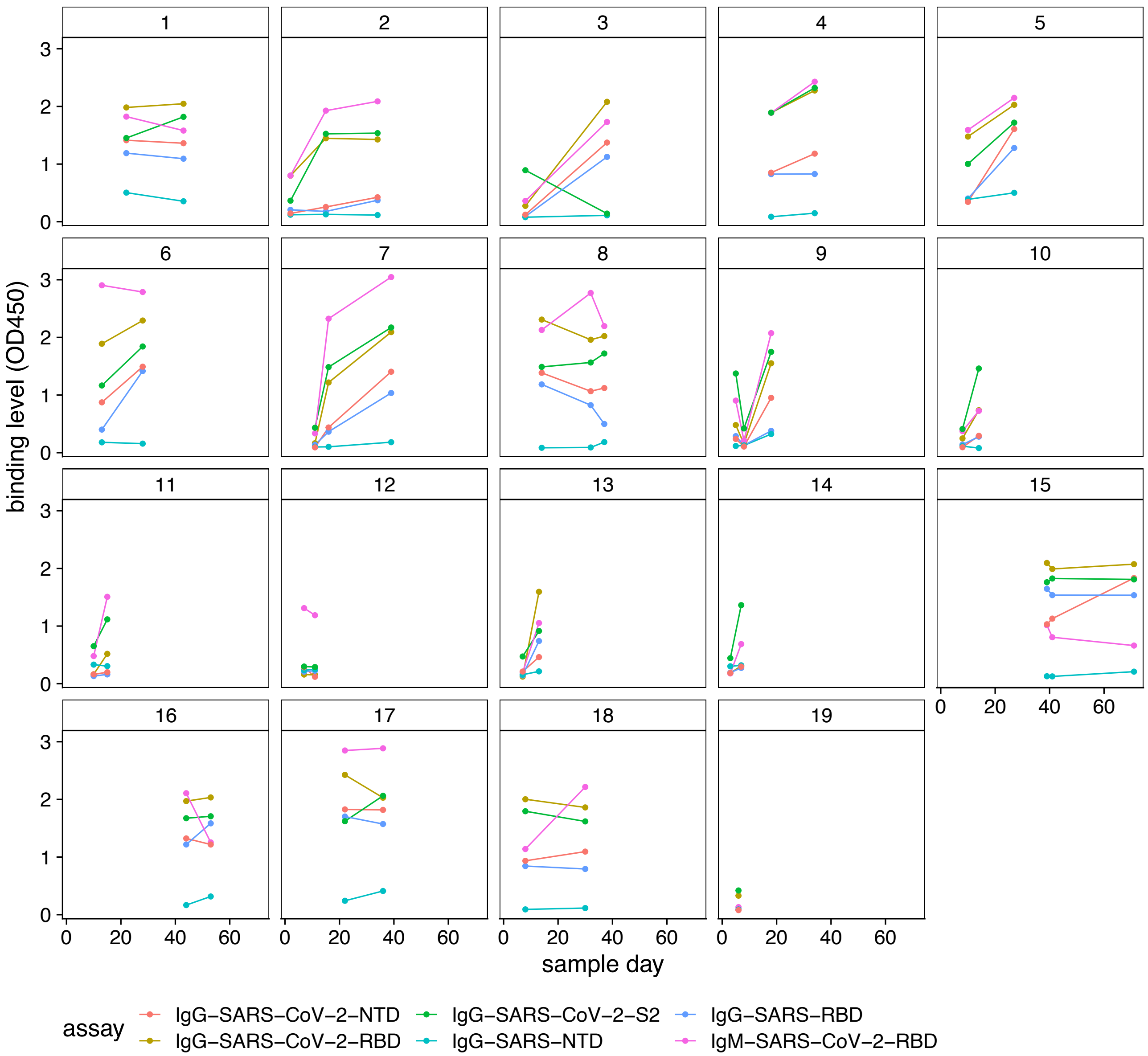

**Figure S5. ELISA binding assays for IgG and IgM repertoires against SARS-CoV-2 and SARS-CoV.** Plasma binding levels (measured by $OD_{450}$ in ELISA) against RBD, NTD, and S2 subdomain of SARS-CoV-2 and against RBD and NTD epitopes of SARS-CoV are shown. As seen in binding assays, many individuals developed a cross-reactive response to SARS-CoV-2 and SARS-CoV. Some individuals showed no increase in IgG binding to SARS-CoV-2 RBD due to already high levels at sampling time or natural variation. For the expansion analysis (Fig. 4), we analyzed only individuals whose IgG repertoires showed an increase in binding to SARS-CoV-2 (RBD): 2, 3, 4, 5, 6, 7, 9, 10, 11, 13, and 14.

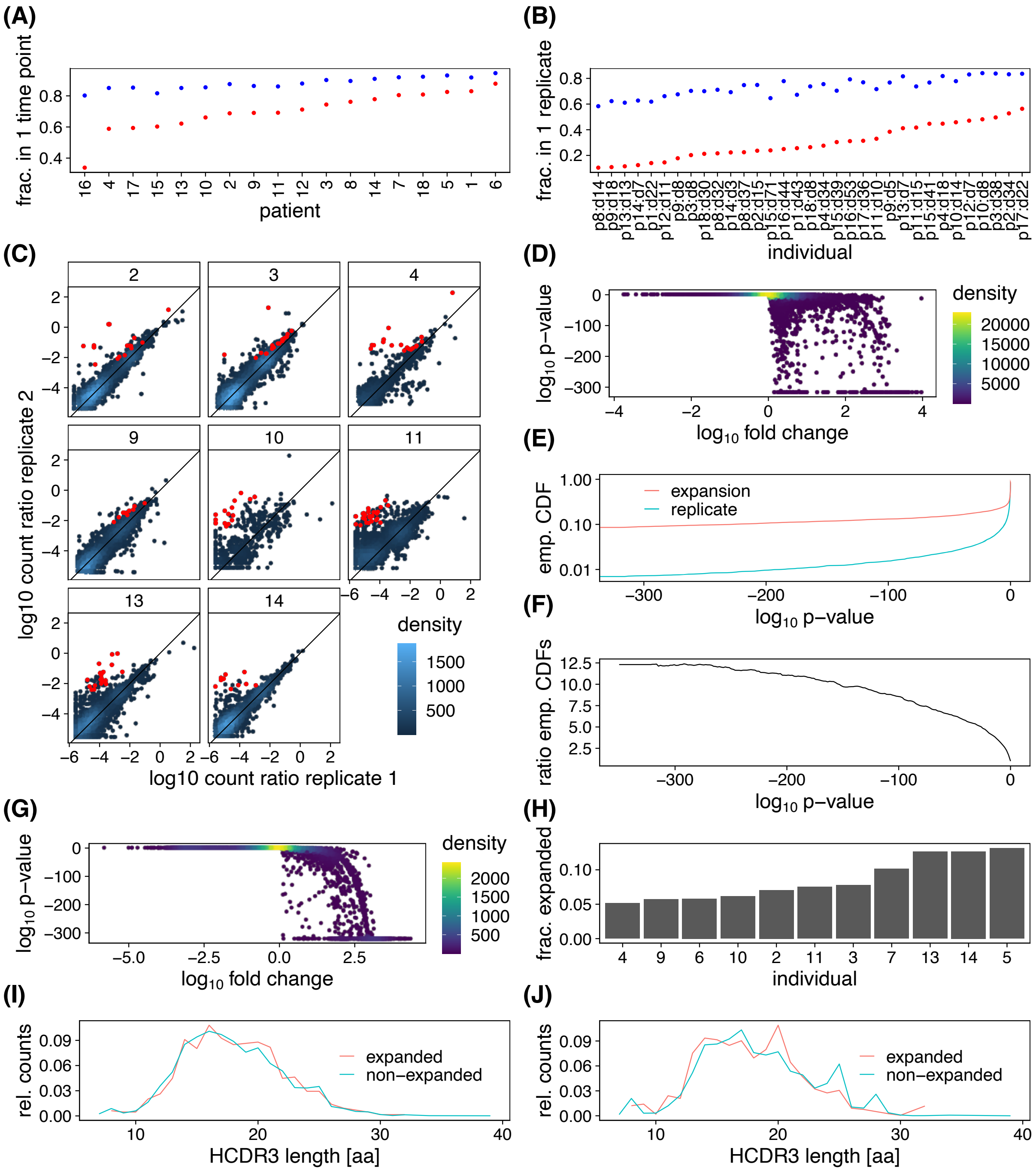

**Figure S6. Expansion supplement.** BCR repertoires are highly under-sampled, and relatively few BCR lineages appear in multiple time points and replicates. **(A)** Fraction of lineages present in only one time point before (blue) and after (red) filtering out small lineages (i.e., those with less than three unique sequences per time point) are shown. **(B)** Fraction of lineages present in only one replicate before (blue) and after (red) filtering out small lineages (i.e., those with less than three unique sequences per time point) are shown. **(C)** The log-ratio of abundance of receptors for all clonal lineages present in two replicates are shown. Each panel shows the test result for a given patient, as indicated in the label. The count density indicates the number of lineages at each point. Lineages that show a significant expansion over time are indicated in red. Since this is replicate data and represents

a null model, red points indicate false positives. **(D)** $\log_{10}$ p-values of the expansion test versus $\log_{10}$ fold change (or odds ratio) for replicate data are shown. Color indicates density of points, and p-values of zero are displayed at the minimum nonzero value. See Methods for normalization, data processing, and hypothesis test. **(E)** Empirical cumulative density functions (CDF) of expansion data from multiple time points (red) and replicate data (blue) show that many more tests in expansion data result in low p-values compared to replicate data. **(F)** Ratio of empirical cumulative density functions (CDF) indicates that at a significance threshold of $10^{-300}$ there are roughly 12.3 times more positives than false positives. **(G)** $\log_{10}$ p-values of the expansion test versus $\log_{10}$ fold change (or odds ratio) for data corresponding to Fig. 4B is shown. Color indicates density of points, and p-values of zero are displayed at the minimum nonzero value. See Methods for normalization, data processing, and hypothesis test. **(H)** Fraction of lineages expanded for different individuals is shown. HCDR3 length distributions of expanded and non-expanded lineages, **(I)** with each lineage having equal weight, and **(J)** with each lineage weighted by the number of unique sequences per time point (excluding singletons) are shown.

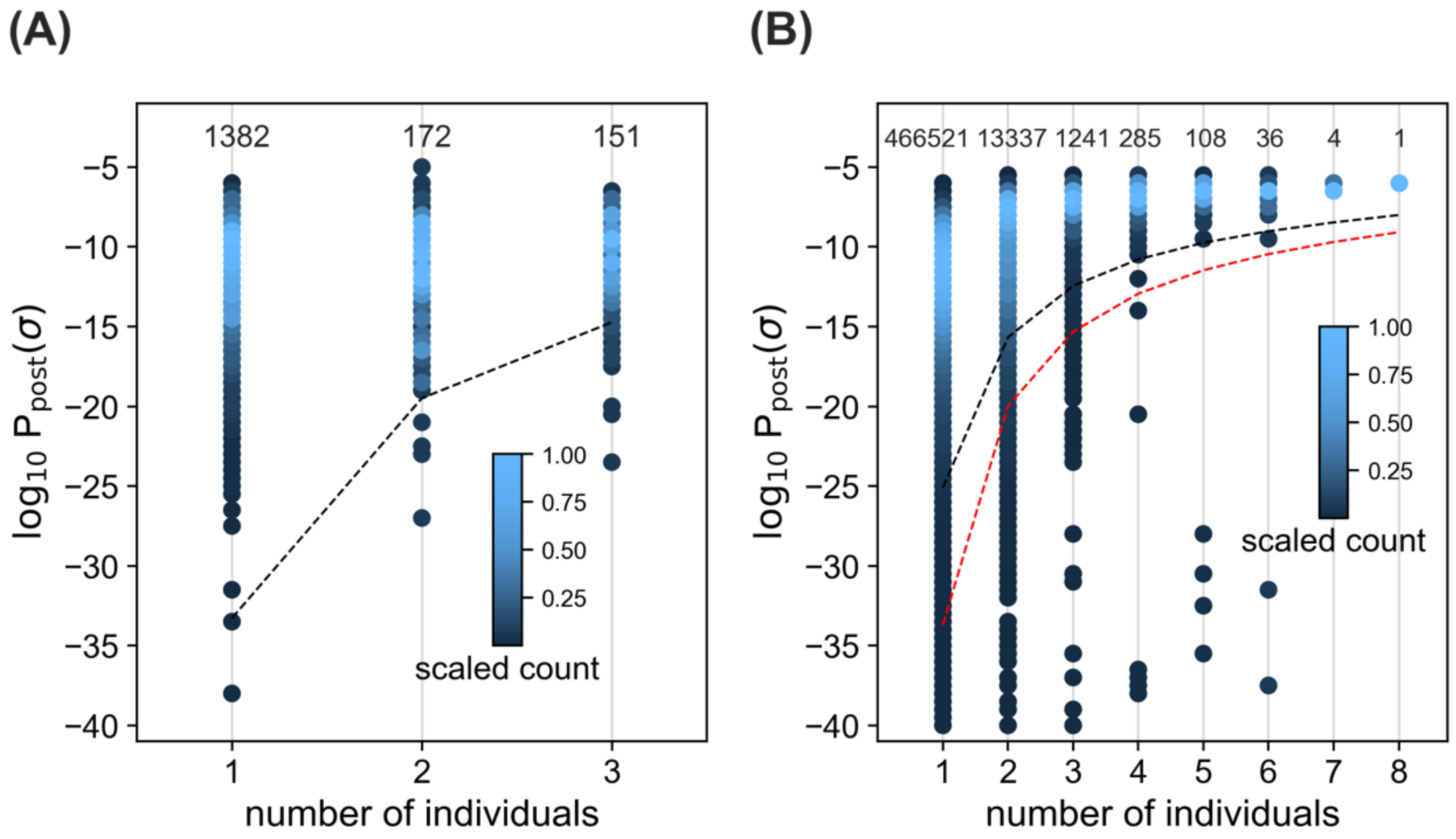

**Figure S7. Sharing of BCRs among healthy individuals. (A)** The density plot shows the distribution of $\log_{10} P_{\text{post}}$ for progenitors of clonal lineages shared in a given number of healthy individuals, indicated on the horizontal axis; histogram bin size is 0.5. The clonal lineages are constructed from the bulk data (Tables S1). The counts in each bin are scaled such that the maximum is equal to one for each column. The numbers above each column indicate the total number of sequences in the respective column. Sharing of rare lineages with $\log_{10} P_{\text{post}}$ below the dashed line is statistically significant (see Methods). **(B)** Similar statistics as in **(A)** are shown but for healthy individuals in the Great Repertoire Project (Briney et al., 2019). Sharing of rare lineages with $\log_{10} P_{\text{post}}$ below the black dashed line is statistically significant (see Methods). For comparison, the dashed line in **(A)** is shown as a red dashed line in **(B)** and extended to eight individuals.

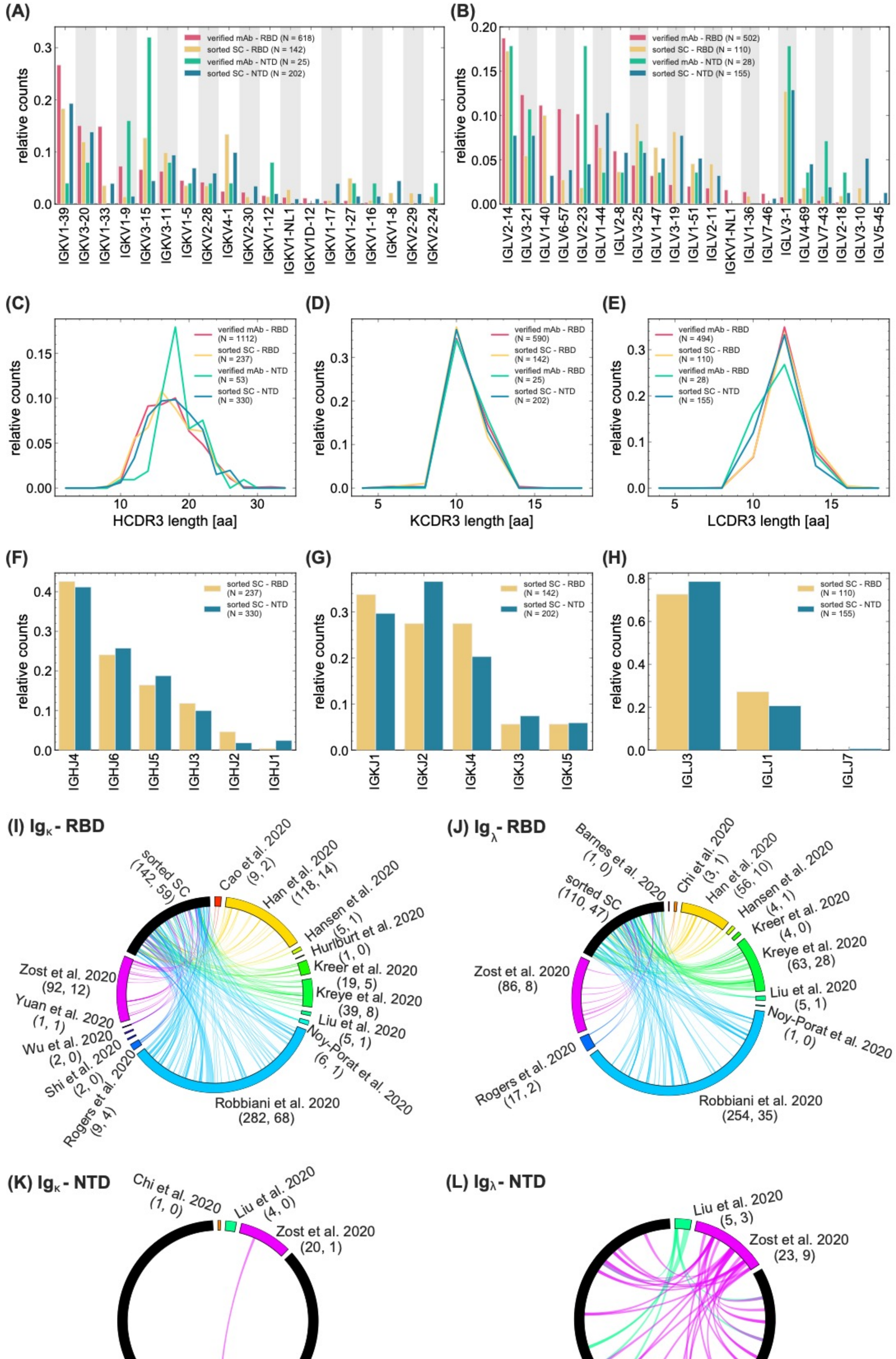

(A)
verified mAb - RBD (N = 618)
sorted SC - RBD (N = 142)
verified mAb - NTD (N = 25)
sorted SC - NTD (N = 202)
relative counts
IGKV1-39
IGKV3-20
IGKV1-33
IGKV1-9
IGKV3-15
IGKV3-11
IGKV1-5
IGKV2-28
IGKV4-1
IGKV2-30
IGKV1-12
IGKV1-NL1
IGKV1D-12
IGKV1-17
IGKV1-27
IGKV1-16
IGKV1-8
IGKV2-29
IGKV2-24
(B)
verified mAb - RBD (N = 502)
sorted SC - RBD (N = 110)
verified mAb - NTD (N = 28)
sorted SC - NTD (N = 155)
relative counts
IGLV2-14
IGLV3-21
IGLV1-40
IGLV6-57
IGLV2-23
IGLV1-44
IGLV2-8
IGLV3-25
IGLV1-47
IGLV3-19
IGLV1-51
IGLV2-11
IGKV1-NL1
IGLV1-36
IGLV7-46
IGLV3-1
IGLV4-69
IGLV7-43
IGLV2-18
IGLV3-10
IGLV5-45
(C)
verified mAb - RBD (N = 1112)
sorted SC - RBD (N = 237)
verified mAb - NTD (N = 53)
sorted SC - NTD (N = 330)
relative counts
HCDR3 length [aa]
(D)
verified mAb - RBD (N = 590)
sorted SC - RBD (N = 142)
verified mAb - RBD (N = 25)
sorted SC - NTD (N = 202)
relative counts
KCDR3 length [aa]
(E)
verified mAb - RBD (N = 494)
sorted SC - RBD (N = 110)
verified mAb - RBD (N = 28)
sorted SC - NTD (N = 155)
relative counts
LCDR3 length [aa]
(F)
sorted SC - RBD (N = 237)
sorted SC - NTD (N = 330)
relative counts
IGHJ4
IGHJ6
IGHJ5
IGHJ3
IGHJ2
IGHJ1
(G)
sorted SC - RBD (N = 142)
sorted SC - NTD (N = 202)
relative counts
IGKJ1
IGKJ2
IGKJ4
IGKJ3
IGKJ5
(H)
sorted SC - RBD (N = 110)
sorted SC - NTD (N = 155)
relative counts
IGLJ3
IGLJ1
IGLJ7
(I) Igκ - RBD
sorted SC (142, 59)
Cao et al. 2020 (9, 2)
Han et al. 2020 (118, 14)
Hansen et al. 2020 (5, 1)
Hurlburt et al. 2020 (1, 0)
Kreer et al. 2020 (19, 5)
Kreye et al. 2020 (39, 8)
Liu et al. 2020 (5, 1)
Noy-Porat et al. 2020 (6, 1)
Robbiani et al. 2020 (282, 68)
Rogers et al. 2020 (9, 4)
Shi et al. 2020 (2, 0)
Wu et al. 2020 (2, 0)
Yuan et al. 2020 (1, 1)
Zost et al. 2020 (92, 12)
(J) Igλ - RBD
Barnes et al. 2020 (1, 0)
sorted SC (110, 47)
Chi et al. 2020 (3, 1)
Han et al. 2020 (56, 10)
Hansen et al. 2020 (4, 1)
Kreer et al. 2020 (4, 0)
Kreye et al. 2020 (63, 28)
Liu et al. 2020 (5, 1)
Noy-Porat et al. 2020 (1, 0)
Robbiani et al. 2020 (254, 35)
Rogers et al. 2020 (17, 2)
Zost et al. 2020 (86, 8)
(K) Igκ - NTD
Chi et al. 2020 (1, 0)
Liu et al. 2020 (4, 0)
Zost et al. 2020 (20, 1)
sorted SC (202, 1)
(L) Igλ - NTD
Liu et al. 2020 (5, 3)
Zost et al. 2020 (23, 9)
sorted SC (155, 22)

**Figure S8. Sequence features of heavy and light chain receptors in sorted single cells and monoclonal antibodies.** The bar graphs show the relative counts for **(A)** the $\kappa -$ chain IGKV-gene usage and **(B)** the $\lambda -$ chain IGLV-gene usage for the verified mAbs reactive to RBD (pink) and NTD (green) epitopes of SARS-CoV-2 (Table S7) and the light chain receptors obtained from the RBD- (yellow) and NTD- (blue) sorted single cell data (Methods). Distributions of the lengths of **(C)** of HCDR (heavy chain), **(D)** KCDR3 ($\kappa -$ chain), and **(E)** LCDR3 ($\lambda -$ chain ) amino acid sequences are shown. **(F)** IGHJ-gene usage, **(G)** IGKJ-gene usage, and **(H)** IGLJ-gene usage of the sorted single cells is shown in relative counts in bar graphs. Colors are consistent between panels and the number of samples used to evaluate the statistics in each panel is indicated in the legend. **(I-L)** Circos plots show matches between the light chain CDR3 sequences of progenitors in the sorted single cell dataset (black) and light chain CDR3 sequences in the verified mAbs (colors) for RBD-reactive **(I)** $\mathrm{IG}_{\kappa}$, and **(J)** $\mathrm{IG}_{\lambda}$ sequences, and for NTD-reactive **(K)** $\mathrm{IG}_{\kappa}$, and **(L)** $\mathrm{IG}_{\lambda}$ sequences. Different colors indicate different studies from which mAbs were pooled. The reference to each study, the total number of mAbs in the study, and the number of mAbs with matching light chain CDR3 to the single cell data are reported in each panel.

TableS1.xlsx

**Table S1. Statistics of BCR repertoire sequence data from healthy individuals, and COVID-19 patients.** The information about individuals in each cohort is shown. Detailed statistics of the processed data for productive BCRs are shown for read abundance, number of singletons, and number of unique sequences for all replicates and sampled time-points in each individual. For each individual, the number of lineages with more than two and ten unique sequences across all time points are shown in separate columns. Read statistics for unproductive receptors pooled from all individuals are shown separately.

| **5'-end primer** | **Sequence (5'-3')** |
|---|---|
| IGHV1 | CCTCAGTGAAGGTCTCCTGCAAGG |
| IGHV2 | TCCTGCGCTGGTGAAACCCACACA |
| IGHV3 | GGTCCCTGAGACTCTCCTGTGCA |
| IGHV4 | TCGGAGACCCTGTCCCTCACCTGC |
| IGHV5 | CAGTCTGGAGCAGAGGTGAAA |
| IGHV6 | CCTGTGCCATCTCCGGGGACAGTG |
| **3'-end primer** | **Sequence (5'-3')** |
| CHG-R | GCGCCTGAGTTCCACGACAC |

**Table S2. List of primers used for PCR amplification of B-cell repertoires samples.**

| Patient (Sex, Age) | Unique Sequences | Lineages (size > 2) | Lineages (size > 10) |
|---|---|---|---|
| 316188 (F, 30) | 204218 | 13138 | 3202 |
| 326650 (F, 18) | 572194 | 28785 | 6139 |
| 326651 (M, 18) | 6174253 | 180993 | 54272 |
| 326713 (F, 25) | 3087960 | 104242 | 29682 |
| 326780 (M, 29) | 297727 | 15195 | 3850 |
| 326780 (M, 29) | 526166 | 26392 | 5039 |
| 326797 (F, 21) | 1538645 | 56487 | 11467 |
| 326907 (F, 29) | 73418 | 4555 | 1124 |
| 327059 (M, 26) | 1932407 | 64786 | 20140 |
| D103 (M, 25) | 169950 | 8423 | 2745 |
| Pooled unproductive | 166704 | 11534 | 1049 |

**Table S3. Statistics of IgG BCR repertoire sequence data from individuals in the Great Repertoire Project.** Because of the massive amount of data provided by the Great Repertoire Project (Briney et al., 2019), only the first three biological replicates were used for each individual to be comparable to the data sampled in this study. Detailed statistics of the processed data for productive BCRs are shown for the number of unique sequences pooled from all replicates for each individual. For each individual, the number of lineages with more than two and ten unique sequences are shown in separate columns. Read statistics for unproductive receptors pooled from all individuals and replicates are shown separately.

## TableS4.xlsx

**Table S4. Statistics of plasma B-cell repertoire sequence data from COVID-19 patients.** The information about individuals in each cohort is shown. Detailed statistics of the processed data for productive BCRs are shown for read abundance, number of singletons, and number of unique sequences for all replicates and sampled time points in each individual. For each individual, the amounts of lineages with more than two and ten unique sequences across all time points are shown in separate columns which are split further by whether the lineages also contained bulk reads.

## TableS5.xlsx

**Table S5. Rare expanding BCRs shared among individuals.** The list of 38 rare progenitors of clonal lineages (i.e., with $P_{\text{post}}$ below the dashed line in Fig. 6) that exhibit lineage expansion in at least one individual is shown. These receptors are indicated by green diamonds in Fig. 6. The presence of these lineages in the plasma B-cell repertoire is indicated in the last column (orange triangles in Fig. 6). The 38 rare expanding lineage progenitors shown here are shared among four to 12 COVID-19 patients.

**TableS6.xlsx**

**Table S6. Single-cell sorted antibodies matched with BCR repertoires of COVID-19 patients.** Receptors from RBD- and NTD-sorted single cell data whose HCDR3 sequences and IGHV-gene clustered with a BCR lineage constructed from the bulk+plasma B-cell repertoires of patients in this study are shown (Methods).

**TableS7.xlsx**

**Table S7. Verified antibodies detected in BCR repertoires of COVID-19 patients.** HCDR3 and IGHV-gene of verified monoclonal antibodies responsive to SARS-CoV-2 (RBD, NTD, and S1) or SARS-CoV-1 epitopes, whose HCDR3 sequences match with a receptor (with up to Hamming distance of one amino acid) in the bulk+plasma B-cell repertoires of patients in this study are shown. Each row indicates a monoclonal antibody family, whose members have similar HCDR3, up to one amino acid difference. Mutations in the repertoire-matched receptors with respect to the original HCDR3 are in red. Single amino acid mutation differences in HCDR3s of monoclonal antibody families are shown in cyan. Patient ID for each repertoire-matched receptor is indicated in the last column. The complete list of verified antibodies is given in Table S8.

**TableS8.xlsx**

**Table S8. Complete list of verified monoclonal antibodies.**